\documentclass[12pt]{article}
\pdfoutput=1

\usepackage{latexsym}
\usepackage{amsthm}
\usepackage{amsmath}
\usepackage[dvips]{graphicx}
\usepackage{amssymb}
\usepackage{multirow}
\usepackage{color}
\usepackage{cancel}

\begin{document}
\baselineskip 0.6cm

\def\simgt{\mathrel{\lower2.5pt\vbox{\lineskip=0pt\baselineskip=0pt
           \hbox{$>$}\hbox{$\sim$}}}}
\def\simlt{\mathrel{\lower2.5pt\vbox{\lineskip=0pt\baselineskip=0pt
           \hbox{$<$}\hbox{$\sim$}}}}

\begin{titlepage}

\begin{flushright}
\end{flushright}

\vskip 2.0cm

\begin{center}

{\LARGE \bf Electroweak Baryogenesis and Dark Matter with an
approximate $R$-symmetry }

\vskip 1.0cm

{\large Piyush Kumar and Eduardo Pont\'on}

\vskip 0.4cm

{\it Department of Physics $\&$ ISCAP   \\ 538 W 120th street \\
Columbia University, New York, NY 10027}

\abstract{ It is well known that $R$-symmetric models dramatically
alleviate the SUSY flavor and CP problems.  We study particular
modifications of existing $R$-symmetric models which share the
solution to the above problems, \emph{and} have interesting
consequences for electroweak baryogenesis and the Dark Matter (DM)
content of the universe.  In particular, we find that it is naturally
possible to have a strongly first-order electroweak phase transition
while simultaneously relaxing the tension with EDM experiments.  The
$R$-symmetry (and its small breaking) implies that the gauginos (and
the neutralino LSP) are pseudo-Dirac fermions, which is relevant for
both baryogenesis and DM. The singlet superpartner of the $U(1)_{Y}$
pseudo-Dirac gaugino plays a prominent role in making the electroweak
phase transition strongly first-order.  The pseudo-Dirac nature of the
LSP allows it to behave similarly to a Dirac particle during
freeze-out, but like a Majorana particle for annihilation today and in
scattering against nuclei, thus being consistent with current
constraints.  Assuming a standard cosmology, it is possible to
simultaneously have a strongly first-order phase transition conducive
to baryogenesis \emph{and} have the LSP provide the full DM relic
abundance, in part of the allowed parameter space.  However, other
possibilities for DM also exist, which are discussed.  It is expected
that upcoming direct DM searches as well as neutrino signals from DM
annihilation in the Sun will be sensitive to this class of models.
Interesting collider and Gravity-wave signals are also briefly
discussed.}

\end{center}
\end{titlepage}

\def\simgt{\mathrel{\lower2.5pt\vbox{\lineskip=0pt\baselineskip=0pt
           \hbox{$>$}\hbox{$\sim$}}}}
\def\simlt{\mathrel{\lower2.5pt\vbox{\lineskip=0pt\baselineskip=0pt
           \hbox{$<$}\hbox{$\sim$}}}}

\renewcommand{\l}{\langle}
\renewcommand{\r}{\rangle}
\newcommand{\be}{\begin{eqnarray}}
\newcommand{\ee}{\end{eqnarray}}

\newcommand{\dd}[2]{\frac{\partial #1}{\partial #2}}
\newcommand{\NN}{\mathcal{N}}
\newcommand{\LL}{\mathcal{L}}
\newcommand{\MM}{\mathcal{M}}
\newcommand{\ZZ}{\mathcal{Z}}
\newcommand{\WW}{\mathcal{W}}

\newcommand{\FO}{{\rm FO}}
\newcommand{\FI}{{\rm FI}}
\newcommand{\TFO}{T_{\rm FO}}
\newcommand{\TFOp}{T_{\textrm{FO}'}}
\newcommand{\YFO}{Y_{\rm FO}}
\newcommand{\zFO}{z_{\rm FO}}
\newcommand{\FOp}{\textrm{FO}'}
\newcommand{\vev}{\textit{vev}}
\newcommand{\vevs}{\textit{vevs}}

\newcommand{\sv}{\langle \sigma v \rangle}

\newcommand{\MPl}{M_{\rm Pl}}

\tableofcontents

\section{Introduction}\label{intro}

Although the Standard-Model is an excellent description of Nature up
to energies of around a hundred GeV, it fails to explain two of the
most important mysteries of our Universe - the nature of Dark Matter
(DM) and the origin of the matter-antimatter asymmetry.~\footnote{We
have nothing to say in this paper about the \emph{biggest} mystery -
the tiny value of the Cosmological Constant.} Many candidates for Dark
Matter - axions, gravitinos, WIMPs, etc.  exist; however the most
popular among them is the WIMP, which can provide the observed relic
abundance from a thermal freeze-out mechanism in a large region of
parameter space.  Moreover, such a particle also arises within many
schemes of beyond-the-Standard-Model physics such as low-scale
supersymmetry, extra dimensions, etc.

Similarly, many popular mechanisms for generating the
matter-antimatter asymmetry exist.  These can be broadly divided into
those which depend on physics at very high scales and those which
depend on physics at the electroweak (EW) scale and below.  Models in
the first class include models of GUT baryogenesis~\cite{Kolb:1983ni}
and leptogenesis~\cite{Fukugita:1986hr}.  On the other hand models in
the second class include those of electroweak
baryogenesis~\cite{Kuzmin:1985mm}, TeV-scale
leptogenesis~\cite{Flanz:1996fb} and Affleck-Dine
baryogenesis~\cite{Affleck:1984fy}.  There also exist models which
propose to explain both the baryon asymmetry of the universe (BAU) and
the DM abundance as an \emph{asymmetry} in a conserved
charge~\cite{Kaplan:1991ah}-\cite{Fujii:2002aj}, \cite{Hooper:2004dc},
although they typically involve a very weakly coupled ``hidden"
sector.

Electroweak baryogenesis is attractive since it only depends on
physics in the visible sector at the electroweak scale; hence the
mechanism is completely testable (at least in principle) at colliders
probing energies of order the electroweak scale, or
astrophysical/cosmological observations probing temperatures (or
times) of order the electroweak scale.  Supersymmetric extensions of
the Standard Model provide an elegant solution to the hierarchy
problem as well as a Dark Matter candidate in the form of the lightest
supersymmetric particle or LSP (assuming $R$-parity conservation).
However, within the minimal supersymmetric Standard Model (MSSM), the
allowed region for a strongly first-order phase transition, which is
one of the important requirements for electroweak baryogenesis, is
severely constrained~\cite{Carena:1996wj} and will soon be tested by
experiments (see \cite{Carena:2008vj} for a recent review).  In
addition, there is tension between having enough CP violation to
produce the required baryon asymmetry and constraints from electric
dipole-moment (EDM) searches~\cite{Pospelov:2005pr}.  It is,
therefore, worthwhile to study extensions of the MSSM with additional
terms in the tree-level scalar potential which could help in making
the transition strongly first-order.\footnote{Alternatively, it has
been argued that new fermions, strongly coupled to the SM, can also
strengthen the EWPT, and lead to a EWBG/DM
connection~\cite{Carena:2004ha}.  Ideas for generating the baryon
asymmetry at the TeV scale \textit{without} a strongly first-order
EWPT have also been proposed~\cite{Shu:2006mm}.} The viability of
singlet extensions of the MSSM with regard to electroweak baryogenesis
has been widely studied \cite{Pietroni:1992in} \cite{Menon:2004wv}.
In particular, Ref.~\cite{Menon:2004wv} studied a singlet extension of
the MSSM called the nMSSM, in which it was shown that it was possible
to have a strongly first-order phase transition as well as generate
the correct relic abundance for the lightest neutralino.  The lightest
neutralino in that case is, however, not completely stable but has a
lifetime longer than the age of the Universe.

In this work, we explore a different extension of the MSSM which can
give rise to electroweak baryogenesis and a WIMP DM candidate.
Assuming standard cosmology, the freeze-out abundance of the WIMP can
account for the full DM abundance in part of the allowed parameter
space.  It is also possible for the WIMP to form an ${\cal O}(1)$
fraction of the DM, or a negligible fraction of DM, in other regions
of the allowed parameter space.  The WIMP DM candidate in the
framework is completely stable.  In addition, the tension with EDM
searches is largely relaxed or inexistent.  A crucial feature of the
model is the existence of an (approximate) $U(1)_{R}$-symmetry,
leading to the following important consequences:

\begin{itemize}
\item The $R$-symmetry leads to new interactions in the scalar
potential which help make the phase transition strongly first-order.

\item Since global symmetries are expected to be violated by
gravitational effects, ``small" $R$-breaking effects are included.
The $R$-symmetry can thus be thought of as an accidental symmetry of
the low-energy theory.  As we will see, the small $R$-breaking helps
in achieving many phenomenologically desirable features at the same
time:\\
a) {\it Viable down-type fermion masses in some realizations (others
do not require $R$-breaking)}.  \\
b) {\it Evade stringent direct-detection bounds from XENON100}.  \\
c) {\it Obtain a relic abundance consistent with observations in part
of parameter space}.

\item The CP-violation required for baryogenesis may have two origins:
a) There are explicit CP phases in the parameters of the scalar
potential, resulting in complex, spacetime-dependent \vevs, inducing
CP-violating phases in the mass matrices for the charginos and
neutralinos, \emph{or} b) All parameters in the scalar potential are real
and CP violation arises from the phases in the $R$-symmetry breaking
Majorana masses of the charginos and neutralinos. In both cases, the
approximate $R$-symmetry suppresses the contribution to EDMs relative
to that in the non $R$-symmetric case.
\end{itemize}

The plan of the paper is as follows.  In Section~\ref{features} we
present the relevant features of the $R$-symmetric scenarios, and
discuss a number of realizations and theoretical considerations.  In
Section~\ref{Potential}, we discuss the zero-temperature properties of
such models, while in Section~\ref{pot-finite} we analyze the
finite-temperature potential and explain the origin of the strongly
first-order phase transition.  In Section~\ref{DM}, we discuss several
aspects related to DM (relic density, direct and indirect searches),
and its connection to the first-order electroweak phase transition
(EWPT).  In Section~\ref{phases} we briefly comment on the generation
of the baryon-asymmetry of the Universe (BAU), and discuss the reasons
that allow to satisfy EDM constraints, even with sfermion masses at
around a TeV. Finally, we consider other signatures in
Section~\ref{other}, and conclude in Section~\ref{conclusions}.  We
also include two appendices with technical details.

\section{Features of the Model}
\label{features}

In this section, we describe the structure of the model at the
electroweak scale in detail.  An $R$-symmetry is well motivated from a
theoretical point of view.  In the global limit, having an exact
$R$-symmetry allows one to construct simple models of dynamical
supersymmetry breaking in accord with the Nelson-Seiberg theorem
\cite{Nelson:1993nf}.  A unique anomaly-free $R$-symmetry also arises
in a superconformal field theory at the superconformal fixed point,
which can be found by $a$-maximization \cite{Intriligator:2003jj}.
One of the first phenomenological models with $R$-symmetry was written
down in \cite{Hall:1990hq}.

Having an $R$-symmetry gives rise to many interesting phenomenological
features. For example, Majorana gaugino masses, trilinear $A$
parameters, and the $B\mu$ parameter are forbidden.  However, gaugino
masses of the Dirac type are allowed.\footnote{Dirac gaugino masses
can also be motivated from ``supersoft" supersymmetry breaking in
which the gauge sector has ${\cal N}=2$ supersymmetry
\cite{Fox:2002bu}.} As a result, this leads to a significant
suppression of flavor and CP-violating effects relative to the MSSM
for ${\cal O}(1)$ flavor-violating soft scalar masses and phases
\cite{Kribs:2007ac}.  Although the minimal $R$-symmetric spectrum does
not give rise to gauge coupling unification, many scenarios for adding
additional matter have been proposed which could help unify the
couplings \cite{Fox:2002bu, Benakli:2010gi, Abel:2011dc, Nelson:2002ca}.

The minimal $R$-symmetric spectrum is shown in Table 1 with the
possible $R$-charge assignments.  The origin of these assignments can
be understood as follows.  The $R$-charges of the adjoints $\{S,T,O\}$
are chosen to ensure that a Dirac gaugino mass term can be written
down with a $D$ SUSY-breaking spurion.  The simplest $R$-charge
assignment of the $Q, U^{c},$ and $H_u$ superfields, so that the
up-type Yukawa couplings are allowed, is given by\footnote{These terms
must be allowed without suppression since the top mass arises from
these.} $R[Q, U^{c}] = 1, R[H_u]=0$.  This leaves the $R$-charges of
the $D^{c}, E^{c}, L$ and $H_d$ superfields.  A number of approaches
can be followed about the $R$-charges of these fields and the
associated couplings:

\begin{table}
\begin{center}
\begin{tabular}{ |c|l|r|r|r|r|}
\hline			
  Superfield & $3\times 2 \times 1$ & \multicolumn{4}{|c|}{$U(1)_R$} \\
\hline
  $Q$ & $(\mathbf{3},\mathbf{2})_{1/6}$ & $1$& $1$&$1$ & $1$\\
  $U^{c}$ & $(\mathbf{\bar{3}},\mathbf{1})_{-2/3}$ & $1$& $1$&$1$ & $1$\\
  $D^{c}$ & $(\mathbf{\bar{3}},\mathbf{1})_{1/3}$ & $-1$ & $1$ & $1$ & $1$\\
  $L$ & $(\mathbf{1},\mathbf{2})_{-1/2}$ & $1$& $1$&$1$ & $1$\\
  $E^{c}$ & $(\mathbf{1},\mathbf{1})_{1}$ & $-1$ & $1$ & $1$ & $1$\\
  $H_u$ & $(\mathbf{1},\mathbf{2})_{1/2}$ & $0$ & $0$ & $0$ & $0$\\
  $H_d$ & $(\mathbf{1},\mathbf{2})_{-1/2}$ & $2$ & $2$ & $0$ & $2$\\
  $S$ & $(\mathbf{1},\mathbf{1})_{0}$ & $0$ & $0$ & $0$ & $0$\\
  $T$ & $(\mathbf{1},\mathbf{3})_{0}$ & $0$ & $0$ & $0$ & $0$\\
  $O$ & $(\mathbf{8},\mathbf{1})_{0}$ & $0$ & $0$ & $0$ & $0$\\
  $R_u$ & $(\mathbf{1},\mathbf{2})_{-1/2}$ & $-$ & $-$ & $2$ & $-$\\
  $R_d$ & $(\mathbf{1},\mathbf{2})_{+1/2}$ & $-$ & $-$ & $2$ & $-$\\
  $H'_u$ & $(\mathbf{1},\mathbf{2})_{+1/2}$ &$-$ & $-$ & $-$ & $2$\\
  $H'_d$ & $(\mathbf{1},\mathbf{2})_{-1/2}$ & $-$ & $-$ & $-$ & $0$\\
\hline  
\end{tabular}
\caption{\footnotesize{Quantum numbers of the superfields in the
model.  The four entries for the $R$-charges of the superfields
$D^{c}$, $E^{c}$, $H_d$ correspond to their values in approaches I-IV
described in the main text.  The fields $\{R_u, R_d\}$ and $\{H'_u,
H'_d\}$ only exist in approaches III and IV, respectively.}}
\end{center}
\end{table}\label{charges}

\begin{itemize}
\item I. One approach to down-type fermion masses was taken in
\cite{Nelson:2002ca} with $R[H_d]=2$, $R[D^{c},E^{c}]=-1$, and
$R[L]=1$.  This allows down-type Yukawa couplings in the
superpotential, at the expense of making the $R$-symmetry anomalous
with respect to $SU(3)_{C}$.  However, since the $R$ symmetry is best
thought of as an accidental symmetry, this does not appear to be an
issue.  There are two ways of generating viable down-type fermion
masses in this scheme.  Small $R$-breaking effects could arise from
Majorana gaugino masses and/or $A$ parameters, which could generate
down-type fermion masses at 1-loop \cite{Dobrescu:2010mk}.
Alternatively, $R$-breaking could arise from a $B\mu$ term, leading to
a small \vev\, for $H_d$ and providing viable down-type fermion masses
via the usual holomorphic Yukawa couplings.

\item II. A model with $R[L,\,D^{c},E^{c}]=1$ and $R[H_d]=2$ is
considered in \cite{Davies:2011mp}, which makes the $R$-symmetry
anomaly free.  In this case, down-type Yukawa couplings are forbidden
in the superpotential, but down-type fermion masses can still be
generated by supersymmetry-breaking terms in the K\"{a}hler potential.
Generating viable fermion masses and an electroweak scale $\mu$-term
then implies that the supersymmetry breaking scale is quite low.
Also, \cite{Davies:2011mp} considered a version of the model without
the singlet $S$.

\item III. Both up-type and down-type fermion masses are allowed if
$R[H_d]=0$, and $R[L,\,D^{c},E^{c}]=1$.  However, terms like $\mu H_u
H_d$ and $\lambda_s S H_u H_d$ which generate Higgsino
masses are forbidden.  Viable Higgsino masses can however be generated
by having additional vector-like doublets $R_u$ and $R_d$ with
$R$-charges $R[R_u,R_d]=2$, which also makes the $R$-symmetry anomaly
free \cite{Kribs:2007ac, Kribs:2010md}.
\end{itemize}

How does our approach compare with those above?  The two main issues
of interest in this paper are the existence of a strongly first-order
electroweak phase transition (EWPT), \emph{and} of a WIMP DM candidate
which could provide all or some of the DM relic abundance and have a
signal in DM experiments.  It turns out that within the $R$-symmetric
setup, the former can be naturally achieved in the presence of the
superpotential term $\lambda_s S H_u H_d$ and the \textit{potential}
term $t_s S + h.c.$ (as discussed in Section~\ref{Potential}).  This
implies that $R[H_d]=2$.  The requirement of a strongly first-order
EWPT, therefore, seems to suggest that we either follow approach I, or
approach II (with the singlet field taken into account).  However,
even with the singlet, approach II requires very low-scale
supersymmetry breaking which typically gives rise to an extremely
light non-WIMP gravitino DM candidate --not desirable from our point
of view.  Approach III does not work within our framework since it
assumes $R[H_d]=0$.

However, it is possible to use an approach which is a combination of
that in II and III, and is consistent with our requirements.  We call
it approach IV and briefly describe it below (we give further details
in Appendix~\ref{IV}).
\begin{itemize}
\item IV. We use the same set of fields as in approach III but focus
on a different region of parameter space.  An appropriate redefinition
of the fields $H_d$, $R_u$ and $R_d$ in approach III can give rise to
viable down-type fermion masses while keeping the $R$-symmetry intact.
The $R$-charges of $L, D^{c}, E^{c}$ are equal to unity, the
same as in approach II and III. However, we redefine the three fields
$H_d$, $R_u$, and $R_d$ with $R[H_d]=0,\,R[R_u]=2,\,R[R_d]=2$ in
approach III in the following way:
\be
\{H_d, R_u, R_d\} \rightarrow \{H'_d,H_d,H'_u\}~,
\ee 
respectively, so that $R[H_d]=2,\,R[H'_u]=2,\,R[H'_d]=0$.  Since
$R[H_d]=2$ now, the coupling $\lambda_s S H_u H_d$ is allowed and a
strongly first order EWPT is possible.  Furthermore, down-type fermion
masses are generated from supersymmetry breaking similar to that in
approach II but with an important difference.  The down-type fermion
masses are only suppressed by the masses of $H'_u$ and $H'_d$, not the
messenger mass $M_{\star}$ as in approach II. Hence, viable down-type
fermion masses \emph{without} $R$-symmetry breaking can be generated
for high supersymmetry breaking scales as long as $M_{H'_u}$ and
$M_{H'_d}$ are only slightly above the superpartner mass scale.  When
the $R$-symmetry is ultimately broken by gravitational effects leading
to a tiny \vev~for $H_d$, there will be an additional
contribution to down-type masses, which will generically be quite
suppressed compared to the one above.
\end{itemize}

Thus, both approach I and IV are consistent with the framework
considered in this paper, and we will consider both possibilities.  We
will show, however, in section \ref{phases} that the two approaches
can in principle lead to different predictions for EDMs.  Finally, let
us comment on the $\mu$-term, which is allowed by the $R$-charges.
Here we assume that a large $\mu$-term (compared to the EW scale) is
absent in the microscopic theory which gives rise to the $R$-symmetric
theory at the electroweak scale.  For example, string selection rules
often exponentially suppress coefficients of terms in the
superpotential which are otherwise allowed by low energy effective
field theory~\cite{Ibanez:2008my}.  Thus, it is possible to naturally
have an EW scale $\mu$ term.\footnote{Any value of $\mu$ is
\emph{technically} natural.} It is then possible to absorb an EW scale
$\mu$ parameter by a redefinition of $S$.  Although the $\mu$-term
appears in the parameters of the scalar potential, since we will
consider the most general form of the potential consistent with the
$R$ symmetry, and with all mass parameters of order the electroweak
scale, there is no loss of generality.

\subsection{Theoretical Details}
\label{susybreaking}

We envision the following simple model of supersymmetry breaking which
preserves an approximate $U(1)_R$, as already discussed in
\cite{Kribs:2010md}.  We will, therefore, be brief and only highlight
the main features.  We assume a hidden supersymmetry breaking sector
with gauge interactions including a $U(1)$ gauge field ${\cal
W}'_{\alpha}$ which gets a nonvanishing $D$-term, and chiral
superfields $X$ charged under the hidden sector gauge group which get
non-vanishing $F$-terms.  $X$ and ${\cal W}'_{\alpha}$ are assumed to
have $R$-charges 2 and 1, respectively.  Supersymmetry breaking is
then mediated to the visible sector by interactions suppressed by the
mediator mass $M_{\star}$.  Since we want to be as model-independent
as possible, we will leave this mass scale unspecified.  Although in
this work we will not be concerned with embedding the model at the
electroweak scale in a UV complete microscopic theory, it is not hard
to envision a variety of ways, which may or may not require a very
small value of $\sqrt{F}$,\footnote{Still with superpartner masses
around the electroweak scale.} in which this can be done.  Requiring a
WIMP DM candidate, however, implies that the supersymmetry breaking
scale be large.  Note that this does \emph{not} automatically imply
gravity or anomaly mediated supersymmetry breaking, as high-scale
gauge mediation models with neutralino LSP can be constructed
\cite{Nomura:2001ub}.

With the above considerations in mind, in addition to the kinetic
terms, the Lagrangian of our model is given by:
\be
\label{Lag}
L &=& L_W + L_{\rm down} + L_{\rm soft}~, \\ 
L_W &=& \int \! d^2 \theta \left({\bf y_u}\,QU^{c}H_u  + \lambda_s\,S H_u H_d + 
\lambda_{T}\, T H_u H_d\right) + {\rm h.c.}~, 
\label{LW} \\
L_{\rm down} &=& \int \! d^2 \theta \left({\bf y_d}\,Q D^{c} H_d + {\bf y_e}\,L E^{c} H_d\right) + {\rm h.c.}~, \hspace{4 cm} {\rm or} 
\label{Ldown} \\
& &  \int \! d^4\theta \left({\bf y_d}\,\frac{{\cal D}^{\alpha}Q\,{\cal D}_{\alpha} D^{c}\,H_u^{\dag}}{M_{H'_d}^2}\,\frac{X^{\dag}X}{M_{\star}^2} + \,{\bf y_e}\,\frac{{\cal D}^{\alpha}L\,{\cal D}_{\alpha} E^{c}\,H_u^{\dag}}{M_{H'_d}^2}\,\frac{X^{\dag}X}{M_{\star}^2}\right) + {\rm h.c.}~, \nonumber\\ 
L_{\rm soft}&=& \sqrt{2} \, c_a \int \! d^2\theta \,\left(\frac{{\cal W}^{\prime \alpha}}{M_{\star}}\right) {\cal W}^{a}_{\alpha}\,\Sigma_a + {\rm h.c.} +
\label{supersoft} \\
& & \left[ c^D_a\,\int \! d^2\theta \,\left(\frac{{\cal W}'^{\alpha}{\cal W}'_{\alpha}}{M_{\star}^2}\right)\,\Sigma_a^2 + {\rm h.c.} \right] + 
c^F_a\,\int \! d^4\theta \,\left(\frac{X^{\dag}X}{M_{\star}^2}\right)\,(\Sigma_a^2 + {\rm h.c.}) + 
\label{BTerms} \\
& &c^F_{ij}\int \! d^4\theta\,\left(\frac{X^{\dag}X}{M_{\star}^2}\right) Q^{\dag}_i\,Q_j~,
\label{SoftMasses}
\ee
where $a=U(1)_Y,SU(2)_L, SU(3)_C$, $\Sigma_a\equiv\{S,T,O\}$, and
$Q_i$ runs over all matter fields.  Note that since there are no gauge
singlets with supersymmetry breaking $F$-terms, the standard trilinear
parameters and the $B\mu$ parameter vanish.  Let us now comment on the
operators above.

The operators in $L_W$ include the holomorphic Yukawa couplings for
the up-quark sector.  Operators in $L_{\rm down}$ are responsible for
down-type fermion masses.  These can either arise from holomorphic
Yukawa couplings (as in approach I) or by supersymmetry breaking (as
in approach IV described above and in Appendix~\ref{IV}).  The
qualitative details of our analysis for the nature of the phase
transition and WIMP DM abundance are essentially \emph{independent} of
the two approaches, hence the results obtained here apply equally well
to both approaches.

Operator (\ref{supersoft}) in $L_{\rm soft}$ not only gives rise to
Dirac gaugino masses, $M_{D_{a}} \equiv c_a\,\frac{\langle
D\rangle}{M_{\star}}$, but also to mass terms for the real
part of the adjoint scalars ${\rm Re}(\Sigma_a)$ as well as
``trilinear" couplings of the form $(M_a \Sigma_a + {\rm
h.c.})\,\left(g_a\sum_i \tilde{q}^{\star}_i\mathbf{T}^a\tilde{q}_i \right)$
\cite{Fox:2002bu}.  These new trilinear couplings will be relevant
when we discuss the EWPT in Section \ref{Potential}.  Operator
(\ref{BTerms}) can be written for fields in real representations of
the gauge group, hence it provides (equal and opposite) mass-squared
terms for the real and imaginary parts of $\Sigma_a$.  Operator
(\ref{SoftMasses}) generates the standard soft mass-squareds for all
scalars.  Note that the adjoint scalars $\Sigma_a$ receive masses from
all three operators above, resulting in different masses for their
real and imaginary parts.

Since the hypercharge adjoint is a singlet $S$ with vanishing
$R$-charge, it is possible to write a trilinear parameter for it:
\be
\label{As}
c_s \int \! d^4\theta\,\left(\frac{X^{\dag}X}{M_{\star}^3}\right) S^3 + 
\left[ c'_s \int \! d^2\theta\,\left(\frac{{\cal W}'_{\alpha}{\cal W}'_{\alpha}}{M_{\star}^3}\right) S^3 + {\rm h.c.} \right]~,
\ee
which is, however, parametrically suppressed relative to other
parameters --it is of ${\cal O}(M_{\rm soft}^2 / M_\star)$ compared to
${\cal O}(M_{\rm soft})$ for others.  In general, there could be an
anomaly-mediated contribution, which is also small compared to other
soft terms.  Higher powers of $S$ appearing in the scalar potential
will be suppressed even further.  So we will neglect the above term
(and other higher order powers of $S$) in our subsequent analysis.

As explained earlier, a small \vev~for $H_d$ is induced once a $B\mu$
term is generated.  Thus, it is natural to have $v_d \ll v_u$ in this
framework.  However, the precise value of the suppression depends on
the mechanism of $R$-breaking.  Note that the usual definition of
``$\tan \beta$" is not very appropriate, and will be discussed in
Section~\ref{phases}.  It is also important to clarify that there is
no light (pseudo) scalar, like the $R$-axion.  This is because the
$R$-symmetry is broken explicitly by small amounts, \emph{and} the
same interactions which break the $R$-symmetry (such as the $B\mu$
term) generate a small \vev~for the $R$-charged $H_d$.  Hence, the
situation is similar to the MSSM with large $\tan \beta$ and small
$B\mu$, in which there is no light pseudo-scalar.  Finally, note that
even in the presence of $R$-symmetry breaking by Majorana gaugino
masses or the $B\mu$ term, a discrete $Z_2$ symmetry (call it
$R$-parity) is preserved making the LSP completely stable.

As has been realized, there also exist potentially dangerous operators
which must be suppressed.  These include:
\be 
& &\int \! d^2\theta \,{\cal W}^{\prime \alpha}{\cal W}_{\alpha Y} + {\rm h.c.}~,
\label{dangerous1} \\
& &\int \! d^2\theta \, {\cal W}'^{\alpha} {\cal W}'_{\alpha} \,S+ {\rm h.c.}~,
\hspace{1cm}
\int \! d^4\theta \,\frac{X^{\dag}X}{M_{\star}}\,S~.
\label{dangerous2}
\ee
Operator (\ref{dangerous1}) induces kinetic mixing between hypercharge
and the hidden $U(1)$, and gives rise to an undesirable large
hypercharge $D$-term, while operators (\ref{dangerous2}) give rise to
tadpole terms for the singlet $S$ which could destabilize the
electroweak hierarchy~\cite{Bagger:1993ji}.  Suppressing the
coefficients of operators arising from the superpotential is
technically natural.  Appealing to that is not possible for the
coefficient of the operator coming from the K\"{a}hler potential,
however.

Many possible mechanisms have been discussed in order to suppress the
coefficients of the above operators.  For example, in gravity-mediated
models or higher-dimensional constructions this may happen if $S$
originates from a a non-abelian gauge theory (in many cases a GUT
multiplet) which is unbroken at a scale not far from the TeV scale
\cite{Kribs:2010md, Chacko:2004mi}.  Within gauge-mediation models, it
is possible to avoid a large tadpole term if the couplings between
adjoints and messengers respect a GUT symmetry \cite{Abel:2011dc}, or
if there is an appropriate parity symmetry forbidding a large tadpole
term \cite{Amigo:2008rc}.  Finally, a large tadpole can be avoided if
the theory itself has a low cutoff $\Lambda_*$, close to the TeV scale
\cite{Kribs:2007ac}.  Since the focus of the paper is not on detailed
model-building, we will be agnostic about the precise embedding of the
model in a UV complete theory and analyze the effective theory defined
by (\ref{LW})-(\ref{SoftMasses}), assuming that the operators in
(\ref{dangerous1})-(\ref{dangerous2}) are sufficiently suppressed,
such that the singlet tadpole is of EW size.  This will be crucial for
the phase transition, as will be seen below.

\section{The $U(1)_{R}$ Symmetric Limit: $T = 0$ Analysis}
\label{Potential}

We are now in a position to study the scalar potential of the model.
With the Lagrangian given in Eq.~(\ref{Lag}) of Section
\ref{susybreaking}, the scalar potential takes the form:
\be
\label{potential}
V &=& V_F + V_D + V_{\rm soft} + V^{(1)}_{\rm loop}~,
\ee
where
\be
V_F &=& \sum_i \left| \frac{\partial W}{\partial \phi_i} \right|^2~,
\hspace{1cm}
V_D = \frac{1}{2}\,\sum_{a=1}^3 (D_2^a)^2 + \frac{1}{2}\, D_Y^2~,
\\
V_{\rm soft} &=& m_{H_u}^2 |H_u|^2 + m_{H_d}^2 |H_d|^2 + m_{s}^2 |S|^2 + 
m_{T}^2 T^{a \dagger} T^{a} \\[0.5em]
&& \mbox{} + B_{T} T^a T^{a} + t_s\,S + B_{s} S^2 + A_s\,S^3 + {\rm h.c}~,
\nonumber
\ee
while $V^{(1)}_{\rm loop}$ refers to the 1-loop contribution to the
effective potential, which will be specified below.  Here $i$ runs
over all fields appearing in the $U(1)_{R}$-symmetric supersymmetric
Standard Model, and $D_Y$ and $D_2^a$ are the hypercharge and
$SU(2)_{L}$ $D$-terms, respectively.  Compared to the MSSM case, the
$D$-terms contain additional pieces associated with the $SU(2)_{L}$
and $U(1)_{Y}$ adjoint fields:
\be\label{D}
D_2^a &=& g (H_u^{\dag}\tau^a H_u + H_d^{\dag}\tau^a H_d + T^{\dag} \lambda^a T) + 
\sqrt{2}\,(M_{D_2}\,T^a + {\rm h.c.})~,\\
D_Y &=& g' (H_u^{\dag} H_u - H_d^{\dag} H_d)  + \sqrt{2}\,(M_{D_1} S + {\rm h.c.})~,
\nonumber
\ee
where $\tau^a$ and $\lambda^a$ are the two and three-dimensional
$SU(2)$ generators respectively.  Note that, as explained in Section
\ref{susybreaking}, the $D$ terms above give rise to new
\textit{trilinear} couplings in the scalar potential, which will be
relevant when we discuss the phase transition.  Also, the masses of
the real and imaginary parts of $S = S_{R} + i \, S_{I}$ and $T =
T_{R} + i \, T_{I}$ are split in Eq.~(\ref{potential}).  For instance,
if $M_{D_{i}}$, $B_{s}$ and $B_{T}$ are real, then $m_{S_{R}}^2 =
m_{s}^2 + 2 B_{s} + 4 M_{D_{1}}^2$, $m_{S_{I}}^2 = m_{s}^2 - 2 B_{s}$,
$m_{T_{R}}^2 = m_{T}^2 + 2 B_{T} + 4 M_{D_{2}}^2$ and $m_{T_{I}}^2 =
m_{T}^2 - 2 B_{T}$.  In this section and the following we will assume
that there are no CP-violating phases in the Higgs scalar potential.
We will comment on the possible presence of such phases, and their
connection to EWBG, in Section~\ref{phases}.  Note also that, within
approach IV, there exist additional fields $H'_u$ and $H'_d$, but
since they are assumed to be parametrically heavier than the other
fields (see Appendix \ref{IV}), their effect on the minimization of
the potential can be neglected .

In order to minimize the above potential, we point out two important
simplifications.  First, EW precision constraints require the triplet
Higgs \vev, $T^{3} \equiv v_{T}$, to be small, which can be achieved
if the triplet soft breaking mass, $m^{2}_{T}$, is in the multi-TeV
range.  Therefore, the effect of the triplet on the minimization of
the potential must be small, and we will set $v_{T} = 0$.~\footnote{We
could keep track of the small triplet \vev, but setting it to zero
will make the physics more transparent.} In addition, we will analyze
the potential in the $R$-symmetric limit.  The $U(1)_{R}$ symmetry
with our $R$-charge assignments implies that $b \equiv B\mu = 0$.
Hence, if $m_{H_{d}}^{2} > 0$, one can easily see that $\langle H_{d}
\rangle = 0$ (i.e.~we avoid spontaneous breaking of the $U(1)_{R}$
symmetry), and the degrees of freedom in $H_{d}$ effectively decouple
from the minimization of the potential.  Therefore, the EW symmetry is
broken by $\langle H_{u}^0 \rangle = v$, with $v \approx 174~{\rm
GeV}$, and the eaten Nambu-Goldstone bosons, as well as the SM-like
Higgs, all arise from $H_{u}$.  In particular, the vacuum is
automatically charge-preserving, exactly as in the SM. Note that when
$R$-symmetry is broken, there will be other terms in the scalar
potential.  However, those terms represent a small perturbation on the
analysis based on the previous approximations, and do not affect the
qualitative results.

The zero-temperature 1-loop contribution to the effective potential,
in the $R$-symmetric limit,\footnote{In approach IV of
Section~\ref{features}, where the $QD^{c}H_{d}$ superpotential term is
forbidden, there are no terms in the one-loop effective potential
involving $H_{d}$.  In approach I, such terms are generated, but since
we will always be in a situation where $H_{d}$ is very small, these
terms play only a minor role in the minimization of the potential.} is
given by $V^{(1)}_{\rm loop} = \frac{1}{2}\,\Delta\lambda\,|H_u|^4$,
where~\cite{Barbieri:1990ja,Carena:1995bx}
\be
\Delta \lambda &\approx& 
\frac{3y_{t}^2}{8 \pi^2} \log \left( \frac{M^{2}_{\rm SUSY}}{m_{t}^{2}} \right) 
\left[ y^{2}_{t} - \frac{1}{4}(g^2 + g^{\prime 2}) \right]~,
\ee
with $y_{t}$ the top Yukawa coupling and $m_{t} \approx 173~{\rm GeV}$
the top quark mass.  Note that we do not write contributions to
$\Delta \lambda$ from the $A$-terms, since these are forbidden by the
$U(1)_{R}$ symmetry, and remain small provided the $U(1)_{R}$
violation is small.  Thus, the size of $\Delta \lambda$ is controlled
by $M_{\rm SUSY}$, which can be taken as the geometric mean of the LH
and RH stop masses.

Minimizing with respect to $H_{u}^0$, we get the condition
\be\label{minimize}
\delta &\equiv& m_{H_u}^2 + \frac{1}{4}(g^2+g'^2+4\,\Delta\lambda)\,v^2 + 
\sqrt{2} \,g' M_{D_1} v_s+ \lambda_s^2 v_s^2 ~=~ 0~,
\label{delta}
\ee
where $v_s$ is the \vev~of $S$.  We will also take advantage of the
fact that $A_{s}$ is expected to be parametrically suppressed with
respect to the other parameters.  So, neglecting $A_{s}$ one gets the
following relatively simple expression for $v_{s}$:
\be
\label{vs}
v_s = -\frac{2\,t_s+ \sqrt{2} \,g' M_{D_1} v^2}{2\,(m_{S_{R}}^2+\lambda_s^2 v^2)}~.
\ee
The vacuum in the above limit thus only depends on 
$\lambda_s$, $m_{S_{R}}^2$, $t_s$ and $M_{D_{1}}$.

The spectrum of Higgses is as follows.  Due to the constraints from EW
precision measurements, the heaviest of the Higgs states are expected
to be the excitations of the triplet $T^{a}$, with a mass of order
$m_{T} \sim \textrm{few TeV}$.  There is also another charged Higgs
and a \textit{complex} neutral scalar, both arising from $H_{d}$, with
masses
\be
m_{H_{d}^0}^{2} &=& m^2_{H_{d}} + \left[ \lambda_{s}^{2} + \lambda_{T}^{2} - 
\frac{1}{4} (g^{2} + g^{\prime 2}) \right] v^{2} - \sqrt{2} g' M_{D_{1}} v_{s} 
+ \lambda_{s}^{2} v_{s}^{2}~,
\label{m2Hd0}
\\[0.5em]
m_{H_{d}^{\pm}}^{2} &=& m_{H_{d}^0}^{2} + \left( \frac{1}{2} \, g^{2} - 
\lambda_{s}^{2} - \lambda_{T}^{2} \right) v^{2}~,
\label{m2Hdmp}
\ee
which could also be somewhat heavy, depending on the soft mass
$m^{2}_{H_{d}}$.

Finally, there is a CP-odd (singlet) scalar with mass $m_{a}^2 =
m_{S_{I}}^2 + \lambda_{s}^2 v^2 - 6 A_{s} v_{s}$, as well as a pair of
CP-even Higgs states that can have couplings to gauge boson pairs.
The latter are linear superpositions of $h_u^0 = {\rm Re}(H_u^0) - v$
and $s = S_{R} - v_{s}$, and the corresponding mass matrix in the
$\{h_u^0, s\}$ basis reads:
\be 
\label{Higgs}
{\cal M}_{H}^2 = \left( \begin{array}{cc} \frac{1}{2}(g^2+g'^2+4\,\Delta\lambda)\,v^2 + \delta & 
v\,[\sqrt{2}g'M_{D_1}+2\lambda_s^{2} v_s] \\ [0.5em]
v \, [\sqrt{2}g'M_{D_1}+2\lambda_s^{2} v_s] & m_{S_{R}}^2
+ \lambda_s^2 v^2 +6A_s v_s
\end{array} \right)~,
\ee
where we included the dependence on $A_{s}$ for completeness, although
we expect it to be negligible.  We also included in the $(1,1)$ entry the
contribution from $\delta$ in Eq.~(\ref{delta}): although it vanishes
in the zero-temperature vacuum, it will give a non-vanishing
contribution at finite temperature, when $v$ and $v_{s}$ are not
identified with their zero-temperature values, and should be included
in the analysis of the next section.  Note that the CP-even states in
(\ref{Higgs}) depend on the same set of microscopic parameters that
determine the vacuum, while the remaining Higgses depend on additional
parameters that do not affect the vacuum (in the limit defined above).
Thus, the stability along the excitations of Eqs.~(\ref{m2Hd0}) and
(\ref{m2Hdmp}), as well as the CP-odd singlet and the triplet
directions, is easily guaranteed by the choice of $m^{2}_{H_{d}}$,
$m^2_{S_{I}}$ and $m^{2}_{T}$, respectively.

The neutralinos and charginos also play an important role in the
collider phenomenology, the DM question, and the EWPT. Since the
neutralinos are Dirac in nature in the $R$-symmetric limit, the mass
matrix in the basis $\{i \tilde{B}, i \tilde{W}^0,
\tilde{H}_d^0~;~\tilde{T},\tilde{S},\tilde{H}_u^0\}$ takes the form:
\be
{\cal M}_{\chi^0} &=& \left( \begin{array}{cc} 0 & {\bf X_N} \\ {\bf X_N}^T & 0\end{array}\right)~,
\nonumber
\ee
where
\be
{\bf X_N}&=&\left( \begin{array}{ccc} 0 & M_{D_1} & m_Z\,s_w\\ M_{D_2}& 0& -m_Z\,c_w\\ 
\lambda_{T}\,v& \lambda_s\,v& \lambda_s\,v_s\end{array}\right)~,
\label{neutralino}
\ee 
and we again set the triplet \vev~to zero.

Similarly, for the charginos we have in the $\{\tilde{T}^-,
i\tilde{W}^-, \tilde{H}_d^-~;~\tilde{T}^+, i\tilde{W}^+,
\tilde{H}_u^+\}$ basis:
\be
{\cal M}_{\chi^{\pm}} &=& \left( \begin{array}{cc} 0 & {\bf X_C} \\ {\bf X_C}^T & 0\end{array}\right)~,
\nonumber
\ee
where
\be
{\bf X_C} &=&  \left( \begin{array}{ccc} 0 & M_{D_2} & 0\\ M_{D_2}& 0& 
\sqrt{2}\,m_W\\ \sqrt{2}\,\lambda_{T}\,v& 0& -\lambda_s\,v_s\end{array}\right)~.
\label{chargino}
\ee
At zero-temperature $v_{s}$ is given by Eq.~(\ref{vs}).  Note that the
neutralino and chargino mass matrices depend on two parameters that do
not enter in the determination of the vacuum: $\lambda_{T}$ and
$M_{D_2}$.  Furthermore, it is easy to find that the three chargino
mass eigenvalues are given by
\be
m_{\chi^{\pm}_{1}} &=& M_{D_{2}}~,
\nonumber \\[0,5em]
m^{2}_{\chi^{\pm}_{2,3}} &=& \frac{1}{2} \left\{ M_{D_{2}}^{2} + (g^{2} + 
2 \lambda_{T}^{2}) v^{2} + \lambda_{s}^{2} v_{s}^{2} \rule{0mm}{8mm} \right.
\\
&& \hspace{5mm} \left. \mbox{}
\pm \sqrt{ \left[ M_{D_{2}}^{2} + (g^{2} + 2 \lambda_{T}^{2}) v^{2} + 
\lambda_{s}^{2} v_{s}^{2} \right]^{2} - 4 \left( \lambda_{s} M_{D_{2}} v_{s} + 
\sqrt{2} g \lambda_{T} v^{2} \right)^{2} } \right\}~.
\ee
We see from the first equation that we must require $M_{D_{2}} \gtrsim
104~{\rm GeV}$ to avoid a chargino lighter than the direct LEP
chargino bound~\cite{CharginoLEP}.  However, we also note that we
cannot take $M_{D_{2}}$ arbitrarily large, since in that limit one
finds $m_{\chi^{\pm}_{2}} \approx \lambda_{s} v_{s} [1 + {\cal
O}(v/M_{D_{1}}, v_{s}/M_{D_{1}})]$, which is typically smaller than
the chargino LEP bound.  Thus, the chargino bound selects a region in
$M_{D_{2}}$ which, together with the region in $M_{D_{1}}$ and
$\lambda_{s}$ preferred by the minimization of the potential (and the
requirement of a strongly first-order phase transition to be discussed
in the next section) leads to a particular composition for the
lightest neutralino, from Eq.~(\ref{neutralino}).  In particular, one
finds a sizable Higgsino component, which has interesting consequences
for the computation of the DM relic density and for direct DM
searches, to be described in Section~\ref{DM}.

\section{A Strongly First-Order EWPT}
\label{pot-finite}

We now turn to the finite-temperature potential, focusing in
particular on highlighting the reasons that lead to a strongly
first-order electroweak phase transition (EWPT).  We assume for
concreteness that the early Universe had a high-enough temperature.
The temperature decreases with the expansion of the Universe, and EWPT
takes place at temperatures of ${\cal O}(T_{EW}) = M_{EW}$, which is
the object of our study in the following.  The main effects arise
already at ``tree-level'', so that it will be sufficient to simply
consider the thermal masses induced by the plasma.  A more refined
analysis based on the full finite-temperature effective potential is
possible, but is not expected to change the conclusions other than in
small details.  This is to be contrasted with the case of the MSSM (or
the SM), where carefully considering up to two-loop (``daisy''
improved) effects~\cite{Espinosa:1996qw} is essential to determine the
strength of the phase transition (see \cite{Carena:2008vj} for a
recent review).

We continue --as in the previous section-- to consider the limit where
the triplet Higgses are heavy so that the triplet \vev~is sufficiently
suppressed to be consistent with the constraints on the
Peskin-Takeuchi $T$-parameter.  Therefore, these particles are
effectively decoupled from the plasma.  However, we will assume that
the four real degrees of freedom in $H_{d}$ and the CP-odd singlet
scalar $S_{I}$ can be considered as light degrees of freedom at the
temperatures of interest (all the masses below $\sim \pi T_{c}$, where
$T_{c}$ is the critical temperature).  The two CP-even Higgs states
from the $h_{u}^{0}$-$s$ system, and the neutralinos/charginos are
also light, hence active in the plasma during the EWPT.

In the $U(1)_{R}$ symmetric limit considered in the previous section,
the minimization of the potential at finite temperature effectively
involves only two degrees of freedom: $\phi \equiv H^{0}_{u}$ and the
singlet scalar, $\phi_{s} \equiv S_{R}$ (at zero temperature, $\langle
\phi \rangle = v$ and $\langle \phi_{s} \rangle = v_{s}$).  The
thermal masses can be read directly from the terms involving $\langle
H^{0}_{u} \rangle = v$ and $\langle S_{R} \rangle = v_{s}$ given in
the previous section.  We use the high-temperature expansion for the
effective potential, so that the leading $T$-dependent terms are
obtained from
\be
V_{T}(\phi,\phi_{s}) &=& \frac{1}{24} \, T^{2} \left[ 2 m^{2}_{H_{d}^{0}} + 
2 m^{2}_{H_{d}^{\pm}} + m^{2}_{a} + {\rm Tr} {\cal M}_{H}^{2} + 6 m^{2}_{W} + 3 m^{2}_{Z} \right]
\nonumber \\[0.5em]
&& \mbox{} + \frac{1}{48} \, T^{2} \left[ 2 \, {\rm Tr} ({\cal M}_{\chi^{0}}^{\dagger} {\cal M}_{\chi^{0}}) 
+ 2 \, {\rm Tr} ({\cal M}_{\chi^{\pm}}^{\dagger} {\cal M}_{\chi^{\pm}}) + 12 m^{2}_{t} \right]
\nonumber \\[0.5em]
&\equiv& c_{\phi} T^{2} \phi^{2} + c_{S^{2}} T^{2} \phi_{s}^{2} + c_{S} T^{2} \phi_{s} + {\rm const.}~,
\ee
where we can drop the field-independent constants, and
\be
c_{\phi} &=& \frac{11}{32} \, g^{2} + \frac{3}{32} \, g^{\prime 2} + \frac{1}{4} \, y^{2}_{t} + 
\frac{1}{8} \, \Delta \lambda + \frac{1}{4} \, \lambda_{s}^{2} + \frac{1}{3} \, \lambda_{T}^{2}~,
\label{cphi} \\[0.5em]
c_{S^{2}} &=& \frac{3}{8} \, \lambda_{s}^{2}~,
\hspace{1.5cm}
c_{S} ~=~ -\frac{1}{4\sqrt{2}} \, g^{\prime} M_{D_{1}}~.
\label{cS}
\ee
We also included the contributions from the electroweak gauge bosons
and the top quark: $m^{2}_{W} = \frac{1}{2} g^{2} \phi^{2}$,
$m^{2}_{Z} = \frac{1}{2} (g^{2} + g^{\prime 2}) \phi^{2}$ and $m_{t} =
y_{t} \phi$ (in the $\tan\beta = \infty$ limit).  The
finite-temperature potential then takes the form
\be
V &=& (\tilde{m}^{2} \phi^{2} + \tilde{\lambda} \phi^{4}) + (2 \tilde{t}_{s} \phi_{s} + 
\tilde{m}_{s}^{2} \phi_{s}^{2} + A_{s} \phi_{s}^{3}) + (2 \tilde{a}_{s} \phi_{s} \phi^{2} + 
\tilde{\lambda}_{s} \phi_{s}^{2} \phi^{2})~,
\label{2dofPotential}
\ee
where the effective coefficients (denoted by a tilde) are given in
terms of the parameters of Eqs.~(\ref{potential}) and
(\ref{cphi})-(\ref{cS}) by
\be
\begin{array}{rclcrclcrcl}
\tilde{m}^{2} &=& m_{H_{u}}^{2} + c_{\phi} T^{2}~,   & &
\tilde{a}_{s} &=& \frac{1}{\sqrt{2}} \, g' M_{D_{1}}~, & &
\tilde{\lambda} &=& \frac{1}{8} (g^{2} + g^{\prime 2}) + \frac{1}{2} \Delta \lambda~,
\\ [0.5em]
\tilde{m}_{s}^{2} &=& m_{S_{R}}^{2} + c_{S^{2}} T^{2}~,  &  &
\tilde{t}_{s} &=& t_{s} + \frac{1}{2} c_{S} T^{2}~, & &
\tilde{\lambda}_{s} &=& \lambda_{s}^{2}~.
\end{array}
\label{OnedofPars}
\ee
The potential of Eq.~(\ref{2dofPotential}) captures the physics to a
good approximation even when allowing for a non-zero but small triplet
\vev, and for small $U(1)_{R}$ violating effects that can induce a
small $H_{d}$ \vev.

We have split the terms in Eq.~(\ref{2dofPotential}) into terms that
depend only on $\phi$, terms that depend only on $\phi_{s}$, and terms
that mix $\phi$ and $\phi_{s}$.  In order to understand the
minimization of this system, and to highlight the underlying mechanism
that leads to a strongly first-order phase transition, it is useful to
start from the situation where the mixing terms vanish.  In addition,
we will make use of the fact that $A_{s}$ is expected to be small [see
comments following Eq.~(\ref{As})], and we will neglect it in the
following.  Then one can see that the potential defines two scales:
$-\tilde{m}^{2}/(2\tilde{\lambda})$ ($ = \phi^{2}$ if $\tilde{m}^{2} <
0$), and $\phi_{s} = - \tilde{t}_{s}/\tilde{m}_{s}^{2}$.  We are
interested in understanding the effects of a non-vanishing singlet
\vev~on $\phi$ once the mixing terms are taken into account.  This can
be understood in a transparent way by writing an effective potential
for $\phi$ after ``integrating out'' $\phi_{s}$ at tree-level (i.e.
using the $\phi_{s}$ EOM).  The mixing terms introduce a further scale
into the problem, $\phi_{\rm tr}^{2} \equiv \tilde{m}_{s}^2 /
\tilde{\lambda}_{s}$, that defines a transition field value for
$\phi$, below which the mixing terms are a small contribution to the
potential compared to the pure singlet terms, and above which they
dominate over the pure singlet terms.  For instance, in the region
where $\phi \ll \phi_{\rm tr}$ one has
\be
\phi_{s} &=& - \frac{\tilde{t}_{s}}{\tilde{m}_{s}^2} \left\{ 1 + 
\left( \frac{\tilde{a}_{s} \tilde{m}_{s}^2}{\tilde{\lambda}_{s} \tilde{t}_{s}} - 1 \right) 
\left[ \frac{\phi^2}{\phi^2_{\rm tr}} + {\cal O}\left( \frac{\phi^2}{\phi^2_{\rm tr}} \right)^2 \right] \right\}~,
\label{phisSmallphi}
\ee
and replacing back in Eq.~(\ref{2dofPotential}) one finds the
effective potential for $\phi$:
\be
V_{\rm eff}(\phi) &=& - \frac{\tilde{t}_{s}^2}{\tilde{m}_{s}^2} + m_{\rm eff}^2 \phi^2 + 
\lambda_{\rm eff} \phi^4 + {\cal O}(\phi^6)
\hspace{1cm}
\textrm{for } \phi \ll \phi_{\rm tr}~,
\label{VeffSmallphi}
\ee
where
\be
m_{\rm eff}^2 &=& \tilde{m}^2 - \frac{2 \tilde{a}_{s} \tilde{t}_{s}}{\tilde{m}_{s}^{2}} + 
\frac{\tilde{\lambda}_{s} \tilde{t}_{s}^{2}}{\tilde{m}_{s}^{4}}~, 
\label{meff}
\\ [0.5em]
\lambda_{\rm eff} &=& \tilde{\lambda} - \frac{1}{\tilde{m}_{s}^{6}} 
\left[ \tilde{\lambda}_{s} \tilde{t}_{s} - \tilde{a}_{s} \tilde{m}_{s}^{2} \right]^{2}~.
\label{lambdaeff}
\ee
We see that one important effect of the singlet \vev~is that it gives
a \textit{negative} contribution to the effective quartic coupling
near the origin.  In fact, as we will see, the condition $\lambda_{\rm
eff} < 0$ characterizes the potentials that exhibit a (strongly)
first-order phase transition at tree-level.\footnote{This point has
also been noted in~\cite{Huber:2006wf}.} The ``instability'' signaled
by the negative effective quartic coupling indicates that there is a
minimum far away from the origin.  In the small field expansion of
Eq.~(\ref{VeffSmallphi}), this ``far away'' minimum can be understood
as arising from balancing the quartic operator against the tower of
higher-dimension operators in Eq.~(\ref{VeffSmallphi}).

That the ``far away'' minimum exists in the above situation is
guaranteed by the fact that (softly broken) supersymmetric potentials
are bounded from below.  One can see this explicitly by considering
the case where $\phi \gg \phi_{\rm tr}$.  In this region, the mixing
terms in Eq.~(\ref{2dofPotential}) dominate over those that depend
only on the singlet, and therefore the singlet \vev~is now given
approximately by
\be
\phi_{s} &=& - \frac{\tilde{a}_{s}}{\tilde{\lambda}_{s}} \left[ 1 + 
{\cal O}\left( \frac{\phi^2_{\rm tr}}{\phi^2} \right) \right]~,
\ee
so that 
\be
V_{\rm eff}(\phi) &=& \textrm{const.} + \left(\tilde{m}^2 - 
\frac{\tilde{a}_{s}}{\tilde{\lambda}_{s}} \right) \phi^2 + \tilde{\lambda} \phi^4 
+ {\cal O}(1/\phi^2)~,
\hspace{1cm}
\textrm{for } \phi \gg \phi_{\rm tr}~.
\label{VeffLargephi}
\ee
Thus, the original (positive) quartic operator controls the
large-field behavior of the potential, while the singlet \vev~affects
only the quadratic and other subdominant terms.

The above arguments suggest that there is a first-order phase
transition, and that it can easily be strong.  Indeed, if $m_{\rm
eff}^2 > 0$ (either at $T=0$ or due to the thermal mass contribution),
then there is a local minimum at the origin, that can be separated by
a barrier from the ``far away'' minimum described above.  For this to
be the case, one must also have that the effective mass, $m^2_{\rm
eff}$, should not be excessively positive, or else it will overwhelm
the effect of the negative quartic operator.  On the other hand, if
this squared mass is negative at $T=0$, the thermal mass contribution
will make it eventually positive, and there will exist a barrier
provided only that $\lambda_{\rm eff} < 0$.  If the symmetry breaking
minimum is the global minimum, then the main effect of increasing the
temperature is to raise the global minimum to be degenerate with the
minimum at the origin at some critical temperature $T_{c}$.  It is
worth emphasizing that this mechanism does not rely on the existence
of a term cubic in $\phi$, and is therefore qualitatively different
from the way the phase transition works in the MSSM (or, had the Higgs
been light enough, in the SM).  As we will see, this allows more
easily for the phase transition to be \textit{strongly} first-order.

Potentials of the form of Eq.~(\ref{2dofPotential}) were 
analyzed in Ref.~\cite{Menon:2004wv}, in the limit that the
temperature dependence of $\tilde{m}_{s}^{2}$ and $\tilde{t}_{s}$ is
neglected, i.e.~setting $\tilde{m}_{s}^{2} = m_{S_{R}}^{2}$ and
$\tilde{t}_{s} = t_{s}$ in Eq.~(\ref{OnedofPars}).  The model studied
in that work was different from ours, and therefore the relation of
the effective parameters to the underlying ones is different in the
two cases.  However, at the level of the effective parameters the two
scenarios are identical, and we simply summarize the main results of
the analysis in~\cite{Menon:2004wv}.  Defining the critical
temperature $T_c$ and the critical \vev~$\phi_c$ ($=\langle \phi
\rangle$ at $T_c$) by:
\be
V(\phi_c, T_c) &=& V(\phi=0,T_c)~,
\\[0.5em]
\left. \frac{\partial V}{\partial \phi} \right|_{\phi=\phi_c} &=& 0~, 
\ee
one finds
\be
\phi_c^2 &=& \frac{1}{\tilde{\lambda}_s} \left(- m_{S_{R}}^2 + \frac{1}{\tilde{\lambda}} \, 
\left| m_{S_{R}} \tilde{a}_s - \frac{\tilde{\lambda}_s t_s}{m_{S_{R}}} \right|\right)~,
\label{phic2} \\
T_c^2 &=& \frac{1}{c_{\phi}} \left[F(\phi_c^2)-F(v^2) \right]~,
\label{Tc2}
\ee
where $c_{\phi}$ was defined in Eq.~(\ref{cphi}), and
\be
F(\phi^2) &\equiv& -2 \tilde{a}_s \phi_{s} - \tilde{\lambda}_s \phi_{s}^2-2 \tilde{\lambda} \, \phi^2~. 
\nonumber
\ee
Here, it is understood that
\be
\phi_s &=& - \frac{2\,t_s + \sqrt{2}\,g'\,M_{D_1} \phi^2}{2\,(m_{S_{R}}^2 + \lambda_s^2 \phi^2)}~,
\nonumber
\ee
which should be compared to Eq.~(\ref{vs}).  Based on the requirement
that $\phi_{c}^{2} > 0$, the authors of Ref.~\cite{Menon:2004wv}
derived a necessary condition for a first-order phase transition that
can be seen to be equivalent to the requirement that $\lambda_{\rm
eff} (T = 0) < 0$, where $\lambda_{\rm eff}$ is the effective quartic
coupling in the small $\phi$ expansion, as defined in
Eq.~(\ref{lambdaeff}).  Our previous comments then give a transparent
interpretation for this condition in regards to the existence of a
barrier separating a minimum at the origin from a ``far away''
minimum.  Furthermore, it clarifies why this is only a necessary --but
not a sufficient-- condition for obtaining two degenerate minima
separated by a barrier, and allows a generalization to the case that
the temperature dependence in $\tilde{m}_{s}^{2}$ and $\tilde{t}_{s}$
is included.  As already mentioned, if the zero-temperature effective
mass parameter, $m^{2}_{\rm eff}$ is positive and excessively large,
it may overwhelm the negative quartic operator in the small $\phi$
region.  What is required is that there exist an ``intermediate''
region where the quartic operator can become relevant: at very small
$\phi$, the potential is determined by the quadratic term, while at
large $\phi$ the higher-dimension operators [in the language of
Eq.~(\ref{VeffSmallphi})] dominate.  For instance, the coefficient of
the $\phi^{6}$ term in Eq.~(\ref{VeffSmallphi}) is given by $G_{\rm
eff} \equiv \tilde{\lambda}_{s} (\tilde{a}_{s} \tilde{m}_{s}^{2} -
\tilde{\lambda}_{s} \tilde{t}_{s})^{2} / \tilde{m}_{s}^{8} > 0$.
Thus, we see that the ``far away'' minimum is of order $\phi_{c} \sim
\sqrt{-\lambda_{\rm eff} / G_{\rm eff}}$, which is obtained by
balancing the $\phi^{4}$ and $\phi^{6}$ operators (the actual value of
the ``far away'' \vev~depends on the full ``tower of operators'' in
Eq.~(\ref{VeffSmallphi}), but the above approximation captures the
qualitative dependence up to an order one coefficient).  Therefore, by
requiring that the quadratic term not be larger than the quartic one
for $\phi_{c} \sim \sqrt{-\lambda_{\rm eff} / G_{\rm eff}}$, one finds
that one must also have $G_{\rm eff} m^{2}_{\rm eff} \lesssim
\lambda_{\rm eff}^{2}$.  Strictly speaking, the above relations
($\lambda_{\rm eff} < 0$ and a not too large $m^{2}_{\rm eff}$) should
hold at the critical temperature, where the existence of the barrier
is crucial.  However, an approximate criterion for a first order phase
transition in terms of microscopic parameters is
\be
\left. \lambda_{\rm eff} \right|_{T=0} &=& \tilde{\lambda} - 
\frac{1}{m_{S_{R}}^{6}} \left[ \tilde{\lambda}_{s} t_{s} - 
\tilde{a}_{s} m_{S_{R}}^{2} \right]^{2} < 0~,
\nonumber \\
\left. G_{\rm eff} \, m^{2}_{\rm eff} \right|_{T=0} &=& 
\frac{\tilde{\lambda}_{s} (\tilde{a}_{s} m_{S_{R}}^{2} - 
\tilde{\lambda}_{s} t_{s})^{2}}{m_{S_{R}}^{8}} \left[ \tilde{m}^2 - 
\frac{2 \tilde{a}_{s} t_{s}}{m_{S_{R}}^{2}} + 
\frac{\tilde{\lambda}_{s} t_{s}^{2}}{m_{S_{R}}^{4}} \right]
~\lesssim~ \left. \lambda_{\rm eff}^{2} \right|_{T=0}~.
\label{EWPTConditions}
\ee
Note that if $\left.  m^{2}_{\rm eff} \right|_{T=0} < 0$, the second
condition is automatic.

The condition that $T_{c}^{2} > 0$ [as obtained from Eq.~(\ref{Tc2})]
was also considered in Ref.~\cite{Menon:2004wv}.  This condition also
has a simple interpretation: it is equivalent to requiring that the
``far away'' minimum be the global one (at $T=0$).  The EW-preserving
minimum has $\phi = 0$ and $\phi_{s} = -t_{s}/m^{2}_{S_{R}}$, which
leads to an associated potential energy $V_{0} = -
t_{s}^{2}/m^{2}_{S_{R}}$ [see Eq.~(\ref{VeffSmallphi})].  At the ``far
away'' minimum one has
\be
V(v,v_{s}) - V_{0} &=& - \frac{v^{4}}{ (1 + \tilde{\lambda}_{s} v^{2} / m^{2}_{S_{R}})} 
\left\{ \lambda_{\rm eff} + 2 \tilde{\lambda} \tilde{\lambda}_{s} 
\left( 1 + \frac{\tilde{\lambda}_{s} v^{2}}{2 m^{2}_{S_{R}}} \right) \frac{v^{2}}{m^{2}_{S_{R}}} \right\}~,
\ee
where $\lambda_{\rm eff}$, as given in Eq.~(\ref{lambdaeff}), is
evaluated at $T=0$ and $v_{s}$ is given in Eq.~(\ref{vs}).  Requiring
that this expression be negative (to ensure that the EWSB minimum is
the global one), results in the condition
\be
\tilde{\lambda} &<& \frac{(\tilde{\lambda}_{s} t_{s} - 
\tilde{a}_{s} m^{2}_{S_{R}})^{2}}{m^{2}_{S_{R}} (m^{2}_{S_{R}} + \tilde{\lambda}_{s} v^{2})^2}~.
\label{VGlobalCond}
\ee
\begin{figure}[t]
\centerline{ \hspace*{-0.5cm}
\includegraphics[width=0.6 \textwidth]{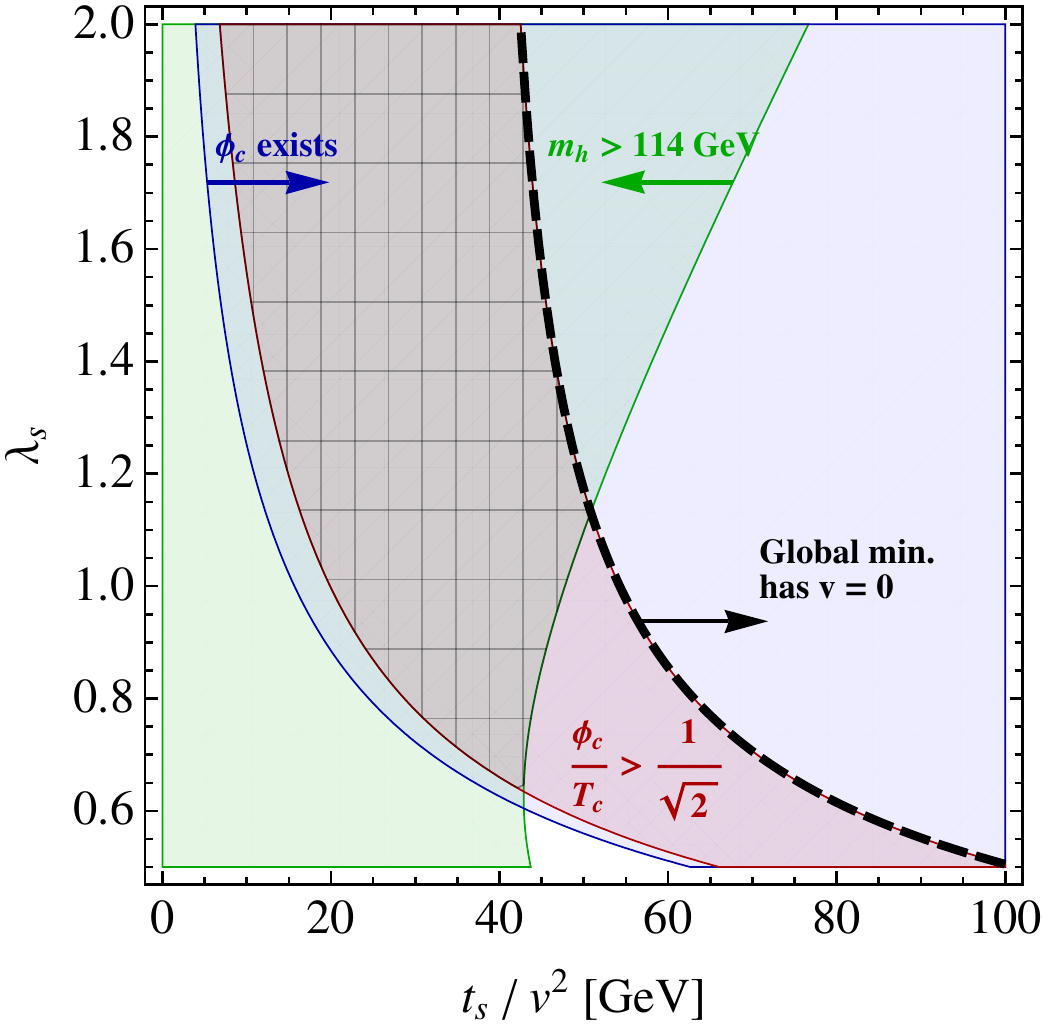}
} 
\caption{\footnotesize{The hatched region in the
$t_{s}$-$\lambda_{s}$ plane leads to a strongly first-order phase
transition (red) and is allowed by the Higgs LEP bound (green).  We
have fixed $M_{D_{1}} = 35~{\rm GeV}$, $m_{S_{R}} = 100~{\rm GeV}$ and
$M_{\rm SUSY} = 2~{\rm TeV}$.}}
\label{fig:ConstraintsEWPT}
\end{figure}
\begin{figure}[t]
\centerline{ \hspace*{-0.5cm}
\includegraphics[width=0.45 \textwidth]{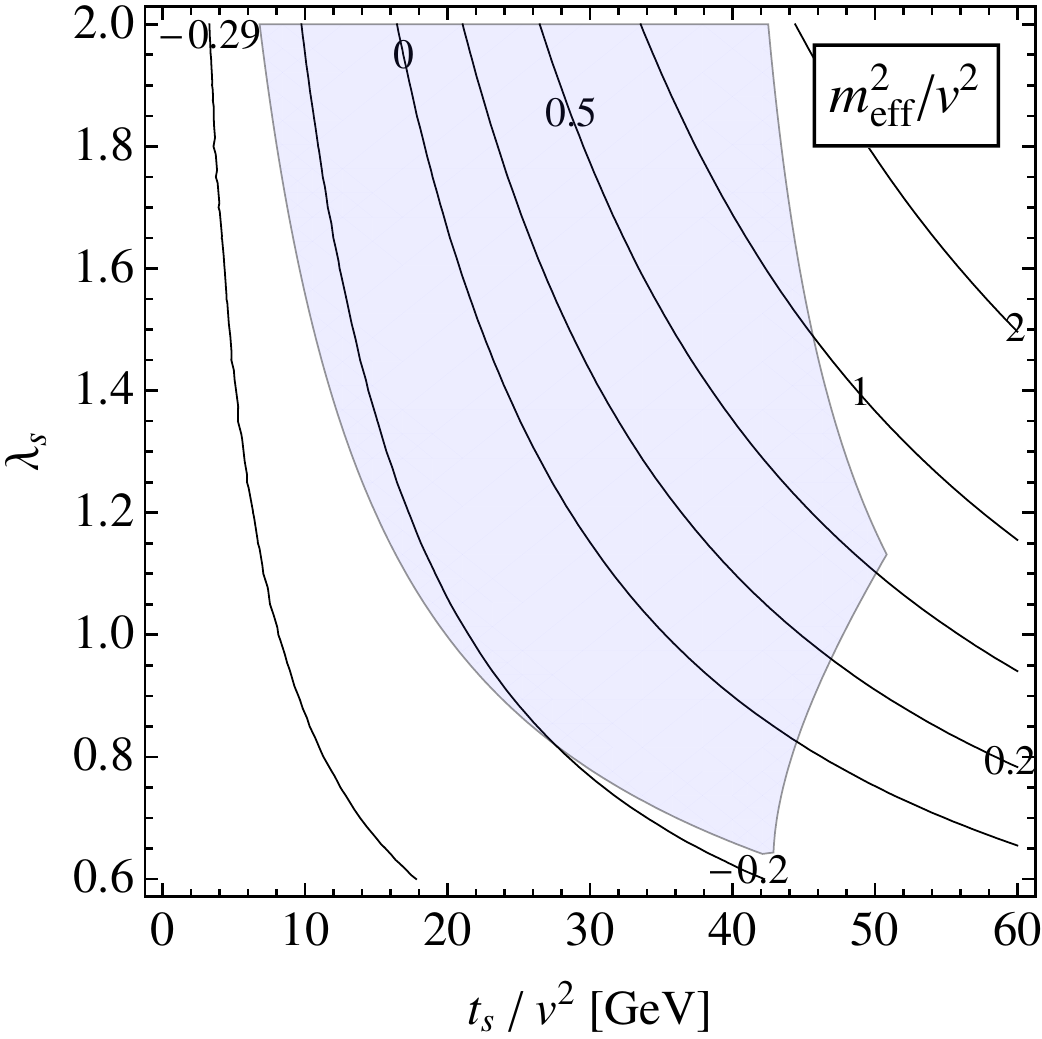}
\hspace*{0.5cm}
\includegraphics[width=0.45 \textwidth]{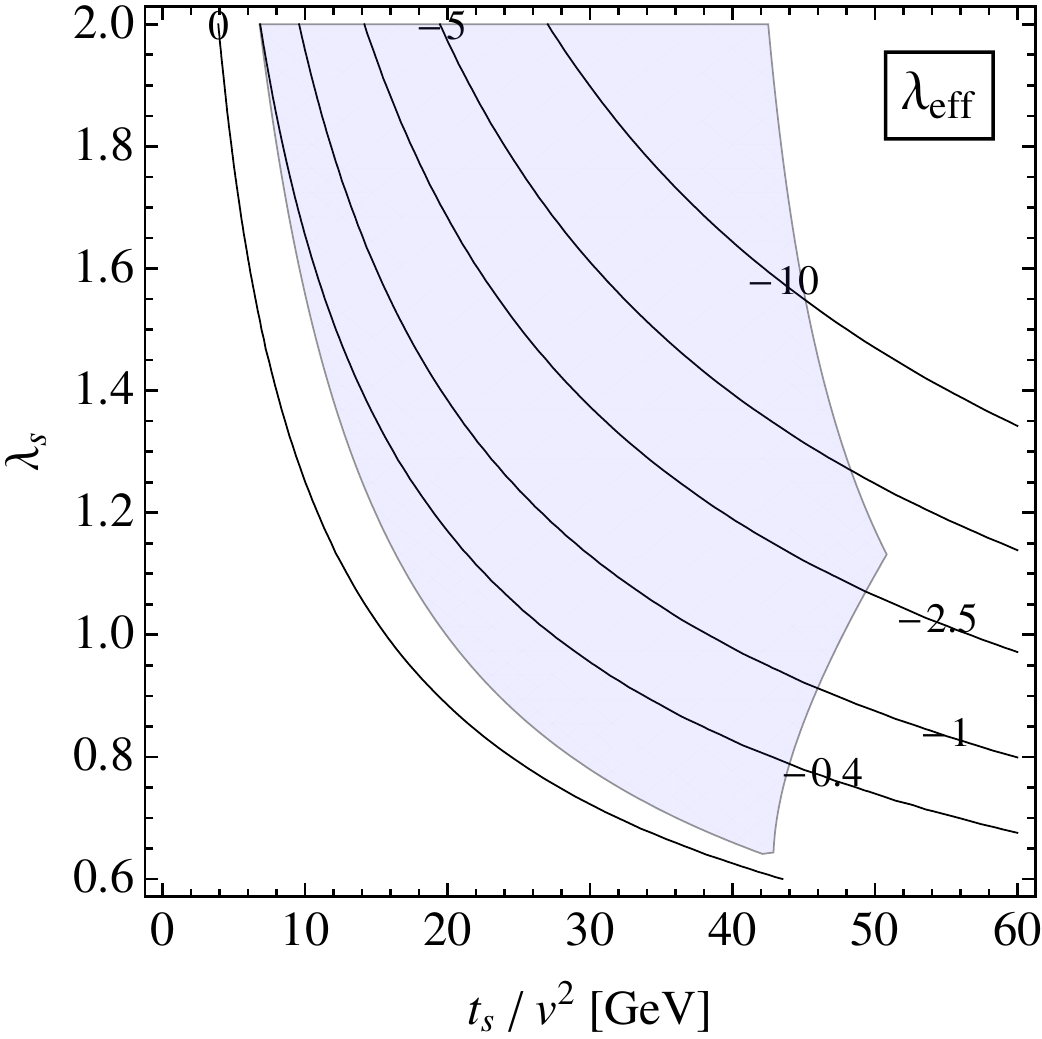}
} 
\caption{\footnotesize{Left panel: lines of constant $m^{2}_{\rm
eff} / v^{2}$, as defined by Eq.~(\ref{meff}).  Right panel: lines of
constant $\lambda_{\rm eff}$, as defined by Eq.~(\ref{lambdaeff}).
Both quantities are evaluated at $T=0$.  The shaded (light blue)
region corresponds to that consistent with a strongly
first-order EWPT and a heavy enough Higgs mass, for the same choice of
parameters of Fig.~\ref{fig:ConstraintsEWPT}}.}
\label{fig:EffectivePars}
\end{figure}
We show in Fig.~\ref{fig:ConstraintsEWPT} how the various constraints
determine the physically interesting region of parameter space.  As
emphasized before, the vacuum structure depends only on four
microscopic parameters: $\lambda_{s}$, $m_{S_{R}}^2$, $t_{s}$ and
$M_{D_{1}}$ (if $A_{s}$ can be neglected).  We show a
projection on the $t_{s}$-$\lambda_{s}$ plane (note that we use the
microscopic $\lambda_{s}$, not the effective $\tilde{\lambda}_{s}$),
for fixed $M_{D_{1}} = 35~{\rm GeV}$ and $m_{S_{R}}^2 = (100~{\rm
GeV})^2$.  In this figure we neglect the temperature dependence in
$\tilde{m}_{s}^{2}$ and $\tilde{t}_{s}$, just as was done in
Ref.~\cite{Menon:2004wv}, but we will make no approximations in the
numerical studies of subsequent sections.  The light blue region (to
the right of the boundary marked as ``$\phi_{c}$ exists'') corresponds
to the condition $\phi_{c}^{2} > 0$, which coincides with the
condition $\left.  \lambda_{\rm eff} \right|_{T=0} < 0$.  The dashed
line corresponds to the boundary where Eq.~(\ref{VGlobalCond}) is
saturated, so that to the right of this line the theory does not break
the EW symmetry.  The interesting region then lies between the two
previous boundaries.  In addition, we denote in a red shade the region
where the EWPT is strongly first-order, as defined by the criterion
$\phi_{c}/T_{c} > 1/\sqrt{2}$ (recall that our normalization is such
that at $T=0$, $\langle \phi \rangle = 174~{\rm GeV}$).  It is clear
from the figure that the condition $\left.  \lambda_{\rm eff}
\right|_{T=0} < 0$ leads almost automatically to a strongly
first-order EWPT. This point has been emphasized more generically
in~\cite{CLP}.  We have also indicated in green the region where the
lightest CP-even Higgs has a mass $m_{h} > 114~{\rm GeV}$, where
\be
m_{h}^{2} &=& \frac{1}{2} \left\{ {\rm Tr} {\cal M}_{H}^{2} - 
\sqrt{({\rm Tr} {\cal M}_{H}^{2})^{2} - 4 \, {\rm Det} {\cal M}_{H}^{2}} \right\}~,
\ee
and ${\cal M}_{H}^{2}$ is given in Eq.~(\ref{Higgs}) (with $\delta =
0$).  Although this condition is overly restrictive, since the
lightest CP-even Higgs state has in general a singlet component, it
illustrates that the Higgs LEP bound is easily consistent with a
strongly first-order EWPT. It should be noted, however, that obtaining
a heavy enough Higgs mass requires that the 1-loop radiative
corrections be significant.  In particular, in the $\tan\beta =
\infty$ limit, the presence of the singlet does not affect the Higgs
mass compared to the MSSM [see the expression for $\tilde{\lambda}$ in
Eq.~(\ref{OnedofPars})].  However, it is also important that there is
no need for either top scalar superpartner to be light in order to
achieve a strongly first-order EWPT, unlike in the light stop scenario
of the MSSM. Hence, both the LH and RH stops can be taken somewhat
heavy.  In the figure, we have chosen $m_{\tilde{t}_{L}} =
m_{\tilde{t}_{R}} = M_{\rm SUSY} = 2~{\rm TeV}$.  The hatched region
corresponds to models that exhibit a strongly first-order EWPT
\textit{and} are allowed by the LEP Higgs bounds.  The shape of the
allowed region is similar for other choices of $M_{D_{1}}$ and
$m_{S_{R}}^2$.  However, as we will see, there are further constraints
coming from the neutralino/chargino sector (in connection to DM) that
prefer relatively small values for $M_{D_{1}}$.  Note also that
$\lambda_{s}$ is required to be sizable although, as we will see, DM
constraints have a preference for the lower range of values.

In Fig.~\ref{fig:EffectivePars} we show again the allowed region, for
the same parameters of Fig.~\ref{fig:ConstraintsEWPT}, and superimpose
the lines of constant $\left.  m^{2}_{\rm eff} / v^{2} \right|_{T=0}$
(left panel) and $\left.  \lambda_{\rm eff} \right|_{T=0}$ (right
panel).  This shows that the allowed region is indeed characterized by
$\left.  \lambda_{\rm eff} \right|_{T=0} < 0$, as well as moderate
$\left.  m^{2}_{\rm eff} \right|_{T=0}$ in the sense that the
effective mass parameter is at most of order $v^{2}$ (it is also clear
that this does not represent an overly restrictive condition).  It can
also be checked that the allowed region satisfies $\left.  G_{\rm eff}
m^{2}_{\rm eff} / \lambda_{\rm eff}^{2} \right|_{T=0} \lesssim 0.9$,
thus validating the physical picture described in the paragraph before
Eq~(\ref{EWPTConditions}).

\section{Pseudo-Dirac Dark Matter - Constraints and Signals}
\label{DM}

We now discuss various aspects of Dark Matter physics in detail.  We
assume that the scale of supersymmetry breaking is high enough that
the lightest $R$-parity odd superpartner, which is a DM candidate, is
a WIMP --usually the lightest neutralino.\footnote{Note that
$R$-parity is still conserved in the model even when $U(1)_R$ is
broken by Majorana gaugino masses and/or the $B\mu$ term.} This is in
contrast to models with low-scale supersymmetry breaking where the
gravitino is likely the DM candidate.  Thus, the relic abundance of
the DM candidate is determined by a freeze-out
calculation,\footnote{Assuming a standard radiation-dominated era after
the end of inflation until after BBN.} and the LSP is also expected
to be seen in direct-detection experiments.  In the $U(1)_R$ symmetric
limit, the neutralino is Dirac in nature and scatters with nuclei
either via $s$-channel squark exchange if it is gaugino like, or via
$t$-channel $Z$-exchange if it has an appreciable Higgsino component,
through a vector-vector interaction which is \emph{not} suppressed by
the DM velocity.  As such, these contributions are subject to
extremely stringent constraints.  Hence, a Dirac neutralino DM
candidate is essentially ruled out.

However, in the presence of (small) $R$-violating operators which are
necessarily present, the two degenerate mass eigenstates of the Dirac
DM split into two Majorana fermions with (slightly) non-degenerate
masses.  Hence the DM in this framework is \emph{pseudo-Dirac} in
nature.  This can naturally allow for consistency with the latest
direct-detection constraints from XENON100~\cite{Aprile:2011hi} if the
splitting between the eigenstates is larger than the typical momentum
transfer (of ${\cal O}(100)$~keV) in a direct-detection experiment,
such that the transition from one state to the other is kinematically
forbidden.  In our framework, we assume that the $R$-breaking is such
that the two Majorana eigenstates are split by $\delta m > 100~{\rm
keV}$ (typically a few GeV, which could arise from anomaly mediation
for example).  Thus, the pseudo-Dirac DM is Majorana-like for the
purposes of direct-detection.  Note that this is quite different from
the scenario of inelastic DM (iDM)~\cite{TuckerSmith:2001hy}, where
the two states are split by an amount comparable to the momentum
transfer in direct detection experiments.  The situation regarding
relic abundance can be completely different, however, since the
freeze-out temperature $T_F$ of the LSP (of ${\cal O}(1\!-\!10)$~GeV
for a WIMP) is much larger than the typical momentum transfer in
direct-detection.  Thus, as long as $\delta m \lesssim T_F$, which can
naturally occur in our framework, the DM properties are similar to
those of a Dirac particle in regards to the relic abundance, as a
result of coannihilation processes being in equilibrium.

\subsection{Relic Abundance}
\label{relic}

As mentioned earlier, we assume for concreteness that the universe
passes through a ``standard" radiation dominated era beginning from a
high temperature due to reheating after inflation, which persists
until the time of matter-radiation equality.  The relic abundance of a
WIMP DM candidate is then determined by a thermal freeze-out
calculation.  The annihilation channels relevant for computing the
relic abundance of DM after freeze-out can be divided broadly into
three classes: a) annihilation into fermions, b) annihilation into
Higgs/$W$/$Z$ states, and c) annihilation into gluon/photon final
states.

What can be said about the above channels within the class of models
considered?  The region of parameter space consistent with a strongly
first-order EWPT and LEP bounds for Higgs and chargino masses (as well
as the direct detection constraints to be discussed in the next
subsection), typically gives rise to a DM candidate with a mass such
that the annihilation channels in $(b)$ above are kinematically
forbidden or suppressed.  Furthermore, since DM is supposed to be
neutral with respect to electric charge and color, DM annihilation
channels in $(c)$ occur via loop effects which may become important
only when the annihilation channels in $(a)$ are suppressed, such as
due to helicity suppression for $s$-wave annihilation of Majorana DM.

Therefore, for the pseudo-Dirac DM with $\delta m \lesssim T_F$,
arising in the scenarios considered in this work, only $s$-wave
annihilation into fermion final states is relevant, which will be
assumed from now on.  This class of annihilation channels can be
further subdivided into three types depending upon the particle
exchanged in the $s$ or $t$ channel.  Thus, the annihilation to
fermion final states ($f\bar{f}$) may proceed through squark/slepton
exchange, $Z$ exchange, or Higgs exchange.  Since the squark/slepton
exchange contribution depends on their (model-dependent) masses and
does not qualitatively affect any other aspect of the physics
considered in this paper, we assume that the masses of
squarks/sleptons are heavy enough such that their contribution to the
relic abundance is subdominant and can be neglected.  Furthermore, the
Higgs exchange contribution is suppressed by the masses of the
fermions,\footnote{The DM is typically not heavy enough to annihilate
into $t\bar{t}$.} hence the dominant $s$-wave annihilation mode is
through $Z$-exchange.  The $Z$-exchange contribution is really a
co-annihilation contribution since $\chi^{0}_1$ and $\chi^{0}_2$ (the
two semi-degenerate Majorana states forming the pseudo-Dirac DM
candidate) co-annihilate through a $Z$ into fermion pairs in an
$s$-wave process.  Also, since only the Higgsino components of
$\chi^{0}_1$ and $\chi^{0}_2$ couple to the $Z$, the relic abundance
of the LSP is correlated with its Higgsino component.  This can be
seen more concretely by considering the pure Dirac ($R$-symmetric)
limit.  The coupling of $Z$ to $\psi_{1}$ (the lightest Dirac
neutralino in four-component notation) is given by:
\be\label{zcoupling}
\left(\frac{\sqrt{g^2+g'^2}}{2}\right)\,Z_{\mu}\,\overline{\psi}_{1} \gamma^{\mu} (c_L P_L+c_R P_R)\,\psi_{1}~,
\ee 
where $c_L =|U^L_{\psi_{1} \tilde{H}_d}|^2$ and $c_R=|U^R_{\psi_{1}
\tilde{H}_u}|^2$, with $U_L$ and $U_R$ the unitary matrices
diagonalizing the Dirac neutralino mass matrix,
Eq.~(\ref{neutralino}), i.e $U_L\,{\bf X_N}\,U_R^{\dag} = {\bf
diag}(m_{\psi_{i}})$.  Thus, $c_L$ and $c_R$ correspond to the
$\tilde{H}_d$ and $\tilde{H}_u$ components of $\psi_1$, respectively.
For a pure Dirac neutralino, the thermally averaged vector-vector
annihilation cross-section to fermions arises from the vector piece of
(\ref{zcoupling}), and is given in the non-relativistic limit by:
%
\be
\langle \sigma\,v\rangle_{Z-{\rm exch}}^{\bar{\psi}_1 \psi_1 \rightarrow\,\bar{f}f} \approx\left( \frac{C^{4}}{2\pi}\right)\,\frac{m_{\psi_1}^2}{(4\,m_{\psi_1}^2-m_Z^2)^2}~,
\label{AnnihilationXS}
\ee 
where we defined
\be
C^4=\left( \frac{\pi^{2}\alpha^2}{c_{W}^4\,s_W^4}\right)\,(c_L+c_R)^2\, \sum_i\, g_{f_i}^{2}~.
\label{C}
\ee 
The sum runs over all SM (Weyl) fermions lighter than $\psi_{1}$
(typically including the bottom, but not the top quark), and $g_{f_i}
= T^{3}_{f_{i}} - 2 s^{2}_{W} Q_{f_{i}}$ is the standard coupling of
$f_{i}$ to the $Z$ gauge boson.  We neglected phase space suppression
factors that are very small even for the bottom quark, for the typical
neutralino masses of interest.  This expression is modified slightly
when $R$-breaking is included, so that the process now corresponds to
co-annihilation of $\chi^{0}_1$ and $\chi^{0}_2$, the masses of which
are split by $\delta m$.  In the limit that $\delta m \ll
m_{\psi_{1}}$, we can write approximately
\be\label{sigmavdelta}
\langle \sigma\,v\rangle_{Z-{\rm exch}}^{\chi_1\chi_2 \rightarrow\,\bar{f}f}&\approx& \langle \sigma\,v\rangle_{Z-{\rm exch}}^{\bar{\psi}_1 \psi_1 \rightarrow\,\bar{f}f}\,\frac{2}{g_{\rm eff}^2}\,(2+\delta)^2\,(1+\delta)^{3/2}\,e^{-x_F\,\delta}~,
\\ 
g_{\rm eff}&=&2+2(1+\delta)^{3/2}\,e^{-x_F\,\delta}~; 
\hspace{5mm}
\delta\equiv \frac{\delta m}{m_{\chi^{0}_1}}~.
\nonumber
\ee 
Here $x_F=m_{\chi^{0}_1}/T_F$ with $m_{\chi^{0}_1} \approx
m_{\psi_{1}}$ and $T_F$ the freeze-out temperature of $\chi^{0}_1$ and
$\chi^{0}_2$, which has to be determined self-consistently.

\begin{figure}[t]
\centerline{ \hspace*{-0.5cm}
\includegraphics[width=0.45 \textwidth]{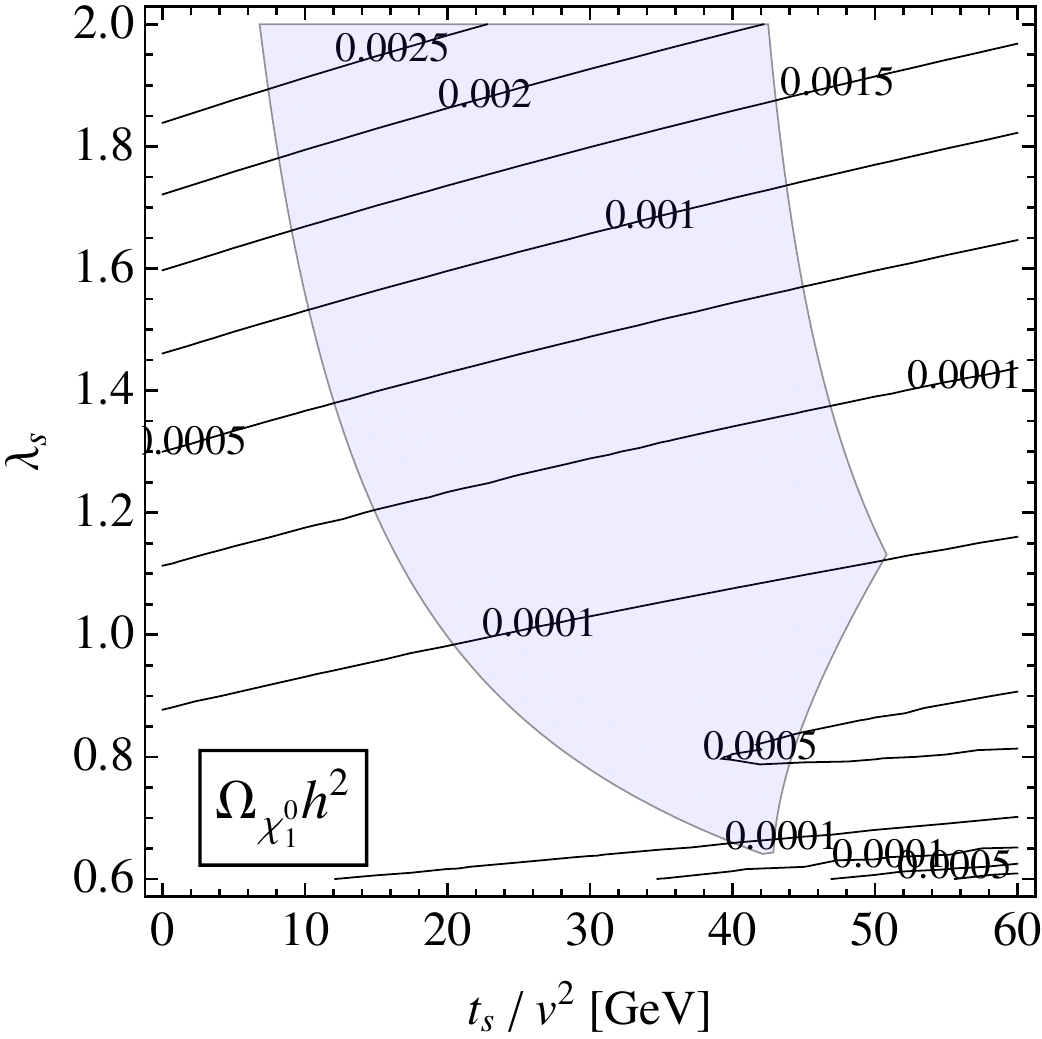}
\hspace*{0.5cm}
\includegraphics[width=0.45 \textwidth]{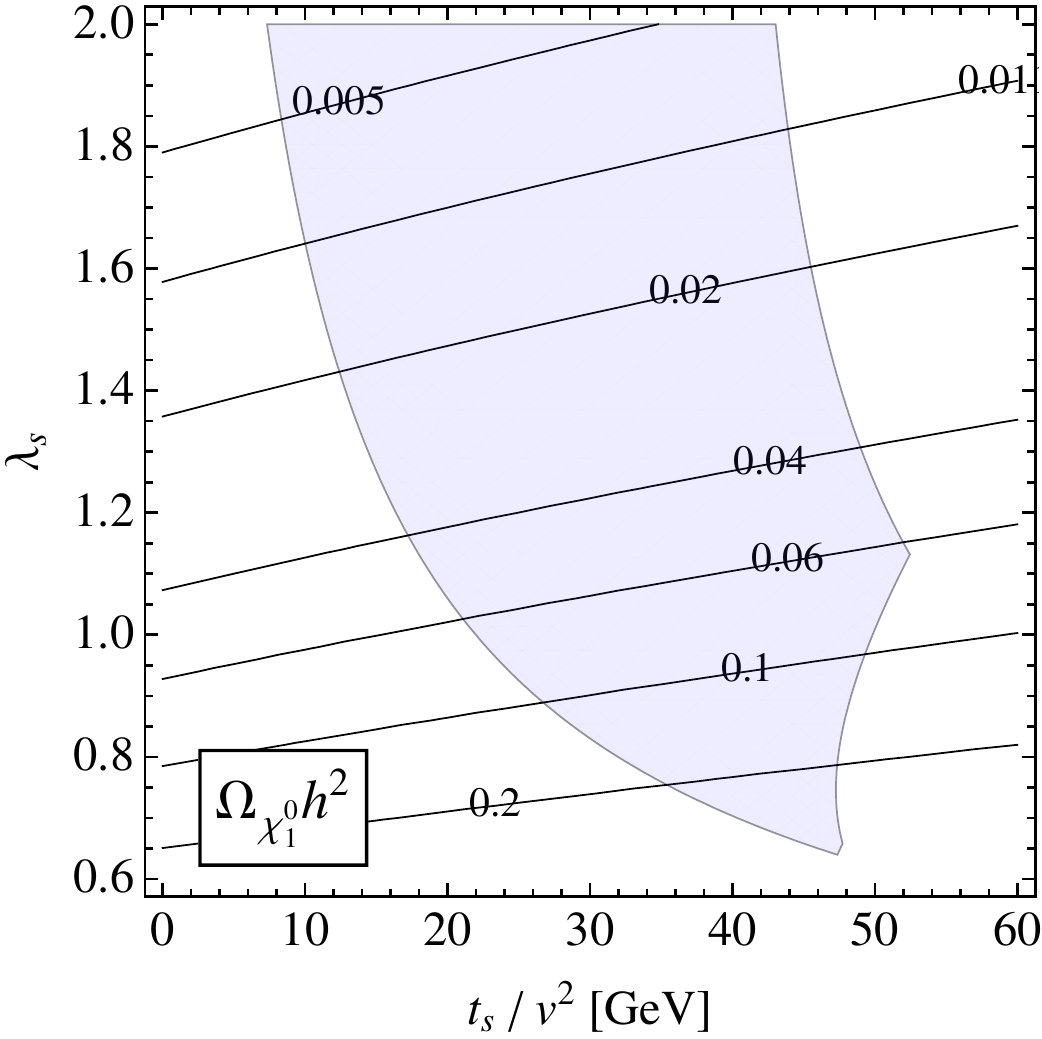}
} 
\caption{\footnotesize{Contours of the LSP relic-abundance $\Omega_{\chi^{0}_{1}}
h^2$ in the $t_s$-$\lambda_s$ plane for the following choice of
parameters.  Left panel: $M_{D_1}$ = 35 GeV, $M_{D_2}$ = -110 GeV,
$m_{S_R}=100$ GeV, $M_{\rm SUSY}$ = 2 TeV; $M_1$ = 5 GeV, $M_2$ = 10
GeV, $b = B\mu = (40\,{\rm GeV})^2$.  Right Panel: $M_{D_1}$ = 60 GeV,
$M_{D_2}$ = -110 GeV, $m_{S_R}=100$ GeV, $M_{\rm SUSY}$ = 2 TeV; $M_1$
= 10 GeV, $M_2$ = 20 GeV, $B\mu = (40\,{\rm GeV})^2$.  The shaded
(light blue) region in the plots corresponds to that consistent with a
strongly first-order EWPT and a heavy enough Higgs mass (in the
$R$-symmetric limit, which should not change much if the small
$R$-violation due to $b$ is also included in the minimization of the
potential).}}
\label{fig:relic-parspace}
\end{figure}
\begin{figure}[t]
\centerline{ \hspace*{-0.5cm}
\includegraphics[width=0.45 \textwidth]{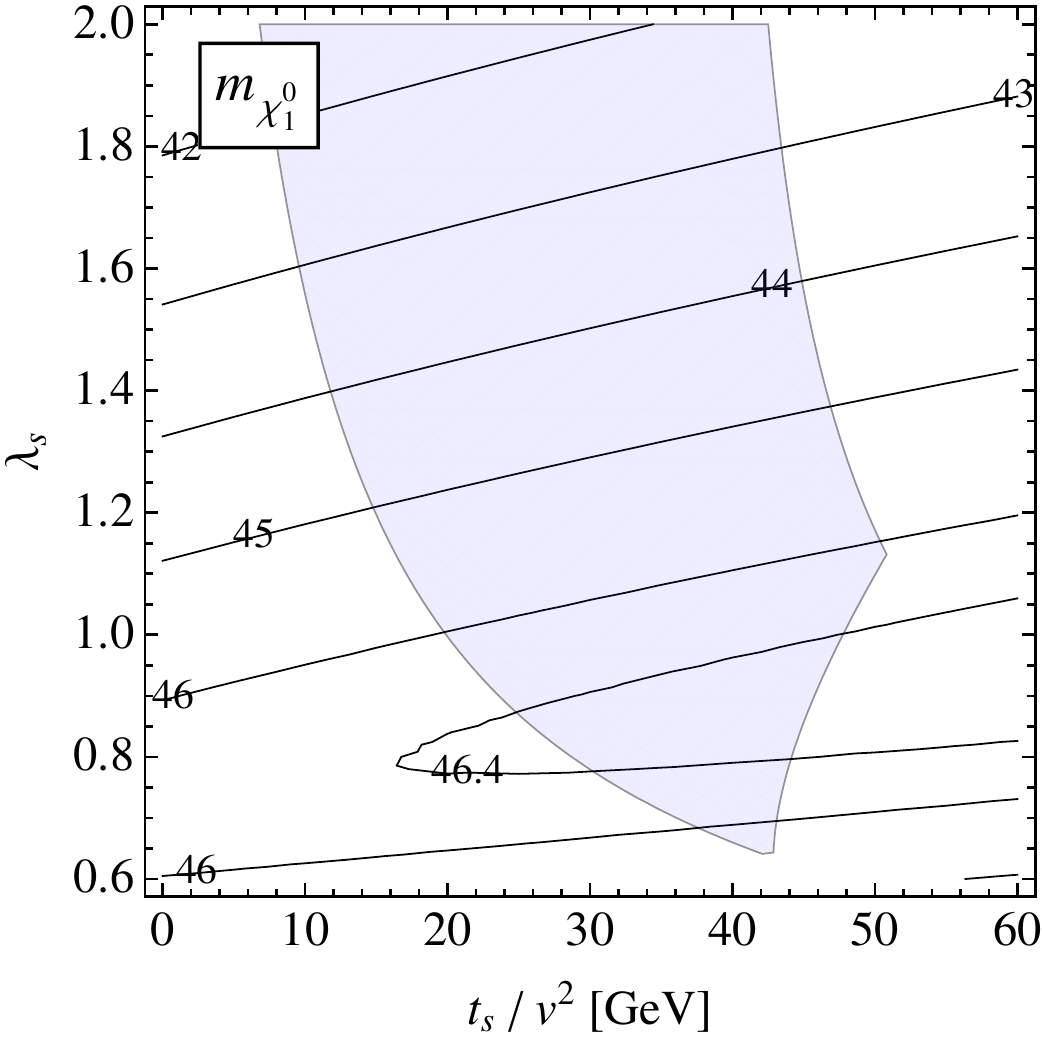}
\hspace*{0.5cm}
\includegraphics[width=0.45 \textwidth]{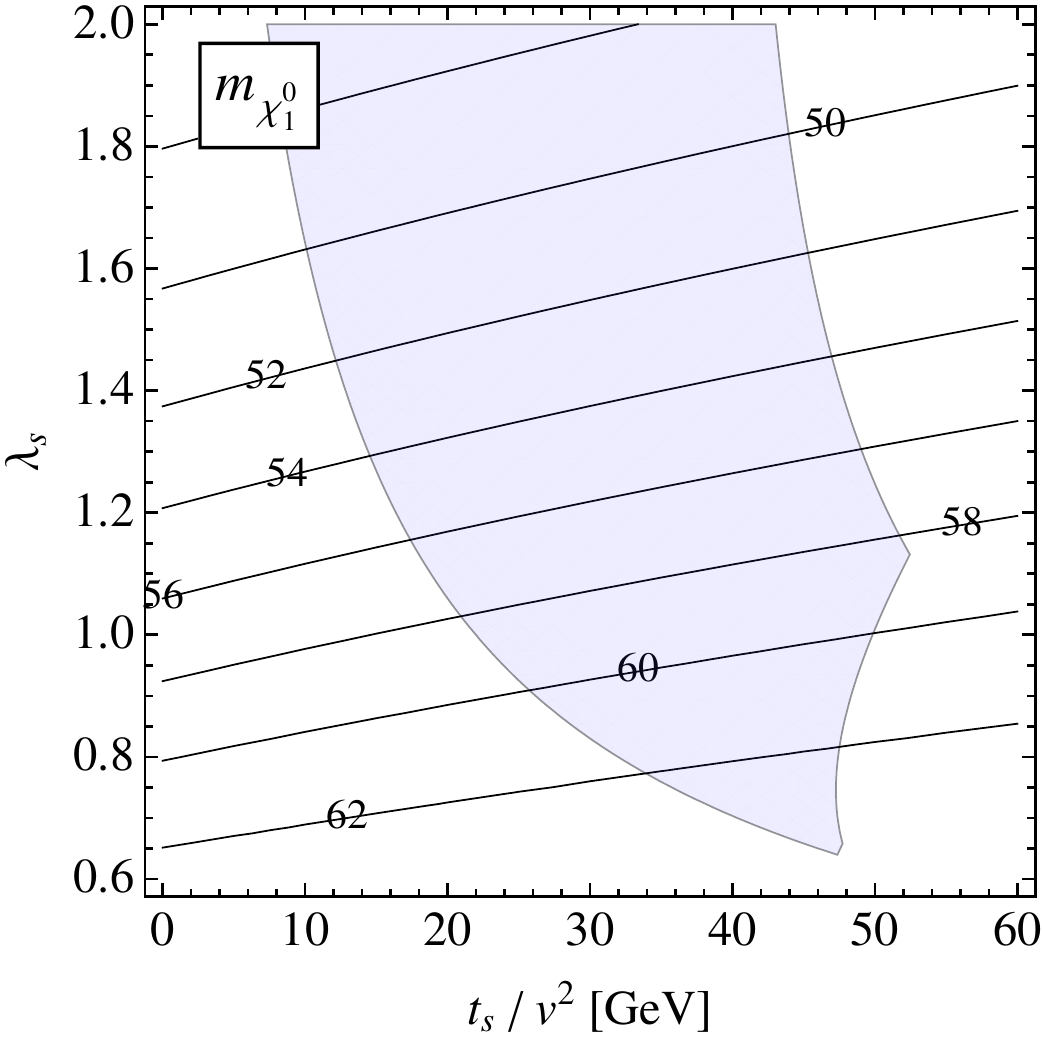}
} 
\caption{\footnotesize{Contours of the lightest neutralino (DM
candidate) mass in the $t_s$-$\lambda_s$ plane for the two choices of
parameters of Fig.~\ref{fig:relic-parspace}}.  The shaded (light blue)
region corresponds to that consistent with a strongly first-order EWPT
and a heavy enough Higgs mass, in the $R$-symmetric limit.}
\label{fig:masses}
\end{figure}
Fig.~\ref{fig:relic-parspace} shows contours of the LSP
relic-abundance for two choices of parameters (see also
Fig.~\ref{fig:parspace-dd} for further motivation for these choices).
The LSP relic abundance in the left panel is always a small fraction
($\lesssim 1 \%$) of the total DM abundance determined by
WMAP~\cite{Komatsu:2010fb}.  However, the LSP relic abundance in the
right panel is typically an ${\cal O}(1)$ fraction of the total DM
abundance, and could even account for the \emph{entire} relic
abundance in a region of parameter space.  The main reasons for the
different relic abundances in the two plots above is the following.
First, the LSP masses for the choice of parameters in the left panel
are close to $m_Z/2$, as seen from the left panel of
Fig.~\ref{fig:masses},\footnote{Fig.~\ref{fig:masses} also illustrates
that the mass of the lightest neutralino (LSP) for these two choices
of parameters is consistent with the LEP bound on the invisible
$Z$-width.  The lightest chargino mass ($\gtrsim$ 105~GeV) for these
choices of parameters is also above the direct LEP bound.} causing a
resonant enhancement in the annihilation cross-section.  No such
resonance enhancement is present in the right panel since the LSP
masses are larger and essentially outside the ``resonance" region (see
right panel of Fig.~\ref{fig:masses}).  Also, the quantity $\delta$ in
the left-panel is generically smaller than that in the right panel.
Although, the precise value of $\delta\equiv
(m_{\chi^{0}_2}-m_{\chi^{0}_1}) / m_{\chi^{0}_1}$ varies throughout
the plots, $\delta$ is roughly given by $M_1/m_{\chi^{0}_1}$, which is
$\sim 1/9$ in the left panel and $\sim 1/6$ in the right panel above.
Since the relic abundance depends exponentially on $\delta$ through
Eq.~(\ref{sigmavdelta}), and $x_{F} \sim 25$, this also has a
non-trivial effect on the relic abundance.  Hence, demanding that the
LSP abundance accounts for all the DM in the Universe favors a
relatively ``large" value of $\delta$ , implying a relatively ``large"
amount of $R$-breaking.  However, since the $R$-breaking is
technically natural,\footnote{In the sense that one has an enhanced
symmetry in the limit when the coefficients of the $R$-breaking
operators vanish.} all effects arising from the breaking of the
$R$-symmetry are naturally suppressed by powers of $\delta$ and loop
factors.  Furthermore, $R$-breaking by Majorana gaugino masses does
not affect the scalar potential and hence the shape of the shaded
(light blue) region in the figures at leading order.

\subsection{Direct Detection}
\label{directDetection}

Having established that the LSP relic density can be a sizable
fraction of the energy content of the universe, we now turn to its
detection.  Since the pseudo-Dirac LSP behaves like a Majorana
particle for direct detection, the two main channels through which it
can scatter off a nucleon is through an axial-vector interaction or a
scalar-scalar interaction.  The axial-vector interaction goes like
$v^2$, where $v$ is the present velocity of DM in the surrounding halo
($\sim 10^{-3}\,c$); hence, the dominant interaction is via a
scalar-scalar interaction, through Higgs exchange.  The Higgs exchange
contribution arises only if the lightest neutralino has a non-trivial
Higgsino component.

\begin{figure}[t]
\centerline{ \hspace*{-0.5cm}
\includegraphics[width=0.45 \textwidth]{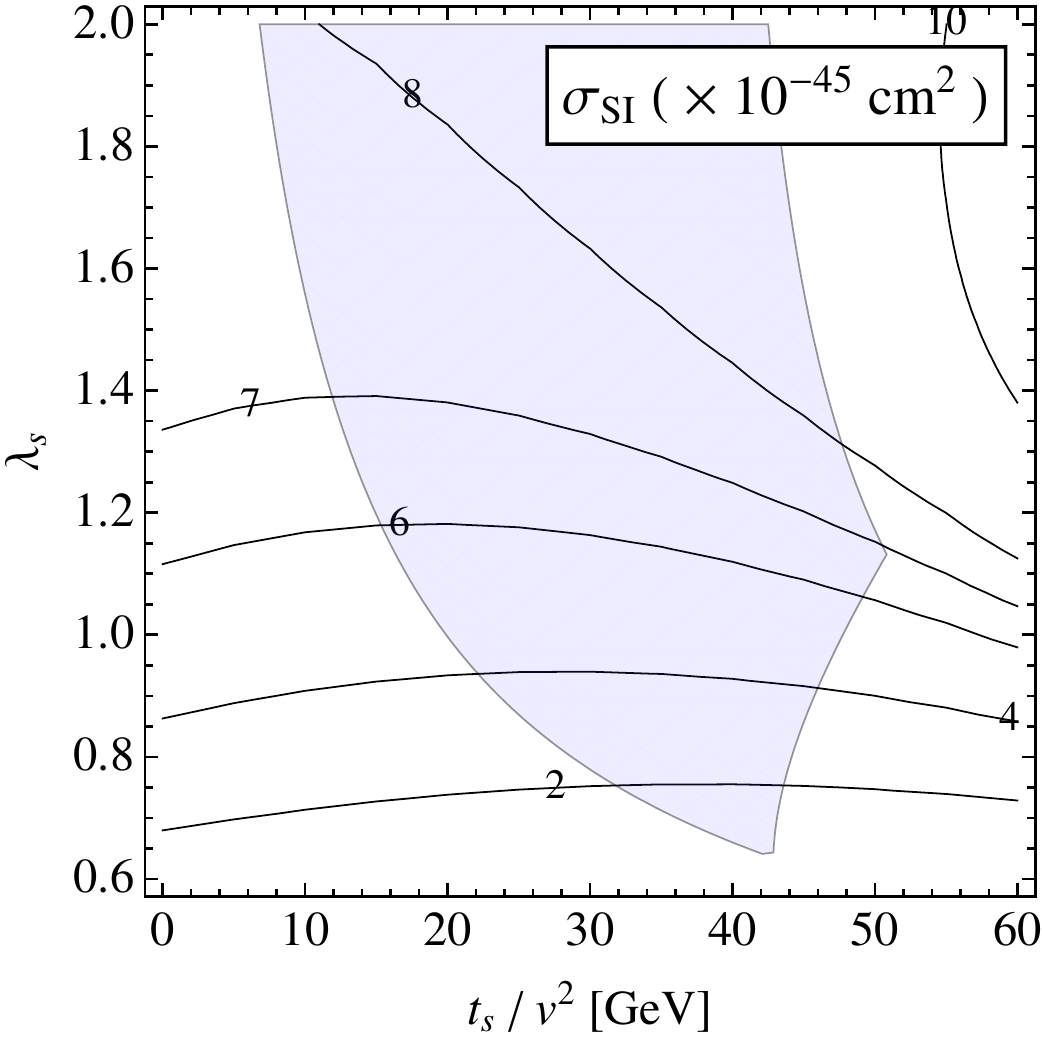}
\hspace*{0.5cm}
\includegraphics[width=0.45 \textwidth]{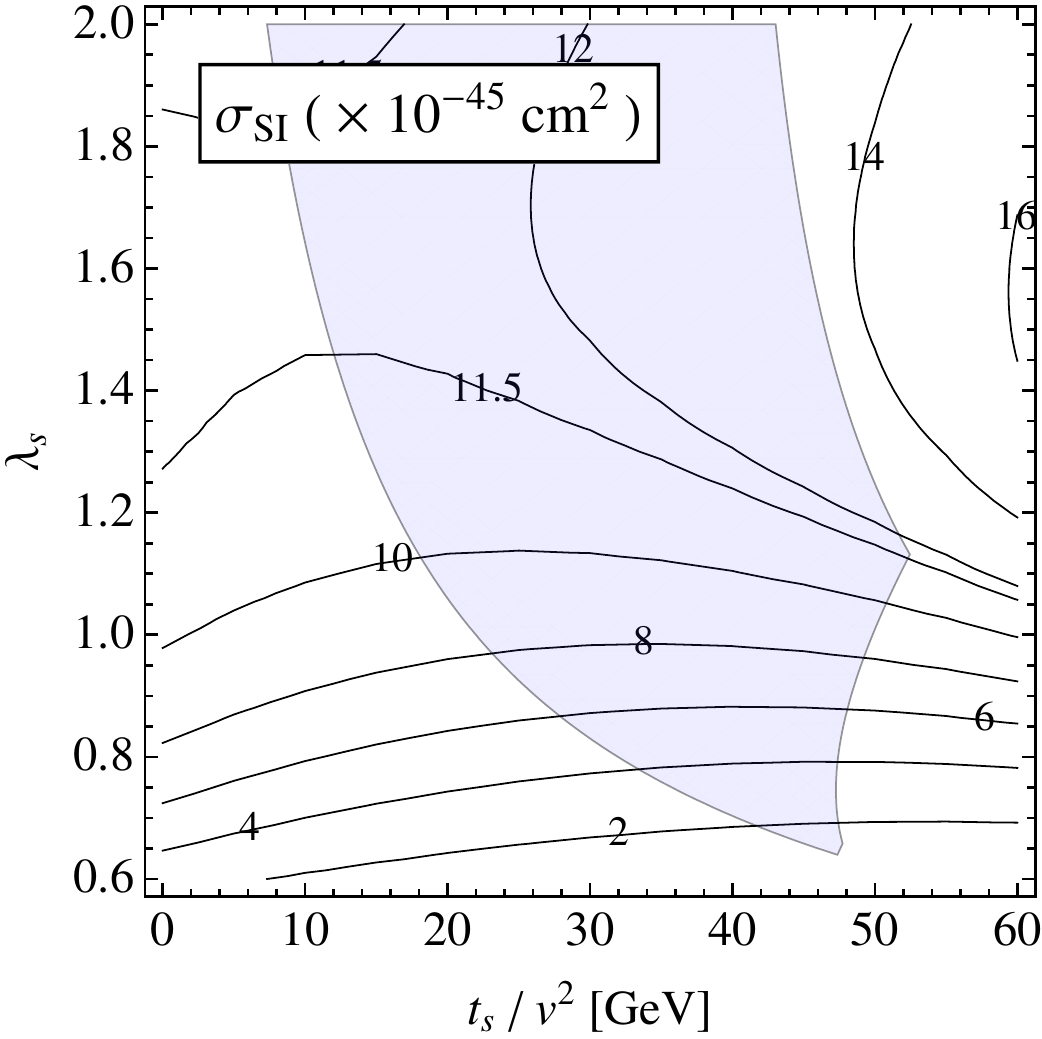}
} 
\caption{\footnotesize{Contours of the spin-independent direct
detection cross-section (in units of $10^{-45}\,{\rm cm}^2$) in the
$t_s$-$\lambda_s$ plane for the following choice of parameters.  Left
panel: $M_{D_1}$ = 35 GeV, $M_{D_2}$ = -110 GeV, $m_{S_R}=100$ GeV,
$M_{\rm SUSY}$ = 2 TeV; $M_1$ = 5 GeV, $M_2$ = 10 GeV, $B\mu =
(40\,{\rm GeV})^2$.  Right Panel: $M_{D_1}$ = 60 GeV, $M_{D_2}$ = -110
GeV, $m_{S_R}=100$ GeV, $M_{\rm SUSY}$ = 2 TeV; $M_1$ = 10 GeV, $M_2$
= 20 GeV, $B\mu = (40\,{\rm GeV})^2$.  The shaded (light blue) region
in the plots corresponds to that consistent with a strongly
first-order EWPT and a heavy enough Higgs mass, in the $R$-symmetric
limit.}}
\label{fig:parspace-dd}
\end{figure}
What is the typical size of the Higgsino component in our framework?
The parameter space for a non-trivial Higgsino component is correlated
with that giving rise to a strongly first-order EWPT in the model.
This can be understood as follows.  Fig.~\ref{fig:ConstraintsEWPT}
shows that the existence of a strongly first-order EWPT with a heavy
enough Higgs mass to satisfy the LEP bounds requires $\lambda_s
\gtrsim 0.6$.  However, it turns out that the coupling of $\chi^{0}_1$
to $h_{u}^{0}$, which is the Higgs state with SM couplings to
nucleons, increases as one increases $\lambda_s$.  Since the
$\chi^{0}_1$-$h_{u}^{0}$ coupling is given by
\be\label{U11}
U_{\chi^{0}_{1},h^{0}_{u}} &=& \left(-\frac{g}{\sqrt{2}}\,U_{\tilde{W}}+\frac{g'}{\sqrt{2}}\,U_{\tilde{B}}\right)\,U_{\tilde{H}_u}+ (\lambda_s\,U_{\tilde{S}}+
\lambda_{T}\,U_{\tilde{T}})\,U_{\tilde{H}_d}~, 
\ee 
where $U_{a}$ corresponds to the $``a"$ component of $\chi^{0}_1$,
$U_{\chi^{0}_{1},h^{0}_{u}}$ is linearly related to the Higgsino
components of the LSP, i.e. $U_{\tilde{H}_u}$ and $U_{\tilde{H}_d}$.
Therefore, the lower bound on $\lambda_s$ for a strongly first-order
EWPT places a lower bound on the Higgsino components of DM, thereby
implying a lower bound on the Higgs-exchange spin-independent
direct-detection cross-section, $\sigma_{\rm SI}$.

Taking into account the exchange of the two CP-even Higgses with
couplings to fermions, the spin-independent nucleon/DM elastic
scattering cross-section can be written, in the non-relativistic
limit, as
\be
\sigma_{\chi_1^{0}}^{{\rm SI}} &\approx&
\frac{4 U_{\chi^{0}_{1},h^{0}_{u}}^{2} g_{\rm hNN}^2}{\pi} \, \frac{
m^2_{N} m_{\chi^{0}_{1}}^2}{(m_{N} + m_{\chi^{0}_{1}})^2} \, \frac{({\bf U}^2_{H,h_{u}} m^2_{h} + {\bf U}^2_{h,h_{u}} m^2_{H})^2}{m^4_{H} m^4_{h}}~,
\label{chiNXS}
\ee
where $h$ and $H$ are the CP-even Higgs mass eigenstates that have an
$h^{0}_{u}$ component [see Eq.~(\ref{Higgs})].  Here ${\bf U}$ denotes
the unitary matrix that diagonalizes the $\{h_u^0, s\}$ system and
${\bf U}^2_{h,h_{u}}$, ${\bf U}^2_{H,h_{u}}$ denote the $h^{0}_{u}$
content in $h$ and $H$, respectively.  We also have that the effective
Higgs-nucleon coupling is $g_{\rm hNN} \approx 0.3 (g m_{N})/(2
m_{W})$, with $m_{N}$ the nucleon mass~\cite{Burgess:2000yq}.  In
Fig.~\ref{fig:parspace-dd} , we show the contours of constant
$\sigma_{\rm SI} = \sigma_{\chi_1^{0}}^{{\rm SI}}$ for the
two different choices of parameters for which we showed the relic
density in the previous section.  Since the Majorana nature of DM is
crucial for direct detection bounds, the set of parameters includes
$R$-breaking Majorana gaugino masses and the $B\mu$ parameter 
(these were also included in the relic density contours).  The
shaded (light blue) region is, however, drawn in the $R$-symmetric
limit for simplicity, as the small $R$-breaking has a minimal effect
on the size and shape of this region.

Fig.~\ref{fig:parspace-dd} shows that a subset of the shaded (light
blue) region is consistent with the latest XENON100 upper bound
($7\times 10^{-45}\,{\rm cm^2}$ for $M_{DM}\approx 50$ GeV) for both
choices of parameters.  The allowed region is smaller in the right
panel of the figure compared to the left one, mainly because $U_{11}$
is larger for the choice of parameters in the right panel compared to
that for the left, and because the lightest Higgs mass is
approximately equal in both plots.  However, it is worth noting that
due to the lower bound on the direct-detection cross-section in the
model, as illustrated in the plots, it is expected that the next round
of results from XENON100 should be sensitive to this class of models,
{\it especially if the LSP accounts for an ${\cal O}(1)$ fraction of
DM}.  If, on the other hand, the LSP accounts for a negligible
fraction of DM, then the detectability of the signal depends on the
product $\rho^{{\rm LSP}}_{\rm local}\,\sigma_{{\rm SI}}$ since it is
this combination which determines the rate at a direct-detection
experiment.\footnote{$\rho^{{\rm LSP}}_{{\rm local}}$ is the density
of the LSP at the Earth's position.}

\medskip

\underline{\textit{Discussion:}} It is important to properly interpret
the plots in Figs.~\ref{fig:relic-parspace} and \ref{fig:parspace-dd}
showing the results for relic-abundance and direct-detection.  The
point to note and appreciate is that depending on where one sits in
the parameter space, the relic abundance and the expected signal for
direct detection can vary considerably.  One possibility, as seen from
the plots in the right panels of Figs.~\ref{fig:relic-parspace} and
\ref{fig:parspace-dd}, is that the LSP of the model accounts for the
entire DM relic abundance and also provides a signal in the next round
of direct-detection experiments.  However, it is perfectly possible
that the LSP abundance is a small fraction of the DM relic abundance
as seen in the left panel of Figs.~\ref{fig:relic-parspace}.  This
would imply that the dominant component of DM has a different origin.
However, a direct-detection signal is nevertheless possible in this
case if $\sigma_{{\rm SI}}$ is large enough (greater than the XENON100
bound of $7\times10^{-45}\,{\rm cm^2}$ for $m_{DM} =50$ GeV) and the
local LSP abundance in the halo is appreciable, since the experiments
are only sensitive to the product $\rho^{{\rm
LSP}}_{local}\,\sigma_{{\rm SI}}$.  Fig.~\ref{fig:parspace-dd} shows
that large SI cross-sections are possible.  Similarly, an
indirect-detection signal in the form of cosmic-ray neutrinos from the
sun for LSPs with a small relic abundance may be possible (see next
subsection).

The LSPs could also provide an ${\cal O}(1)$ fraction of DM, so that
DM consists of more than one non-negligible components.  For example,
axions provide an elegant solution to the strong CP-problem and also
provide a good candidate for DM. They also naturally arise within
string theory.  Hence, it is a natural possibility that both WIMPs and
axions provide an ${\cal O}(1)$ contribution to the total DM in the
Universe.  In fact, it has been shown that this can occur in
well-motivated particle physics frameworks arising from string
theory~\cite{Bobkov:2010rf, Acharya:2010zx}.  Finally, the plot in the
right panel of Fig.~\ref{fig:relic-parspace} shows that it is also
possible for the LSP abundance to be {\it larger} than the total relic
abundance.  However, the implicit assumption in these plots is that
the contribution from the annihilation channels mediated by
$t$-channel squark/slepton exchange is negligible.  The above
assumption is relaxed if the squarks/sleptons are light enough, so
that these regions may also become viable in that situation (as we
will discuss in Section~\ref{phases}, light sfermions can be
consistent with EWBG and bounds on EDMs).

Thus, even if the above framework provides a correct description of
electroweak scale physics, the precise region of parameter space
selected by Nature can only be determined by detailed measurements
from a combination of experiments and observations in particle physics
and astrophysics.

\subsection{Indirect Detection}

In addition to ``directly" detecting the WIMP DM candidate by its
recoil when it scatters with nuclei in a detector, another possible
observable signal of WIMPs arises from the production of electrons,
positrons, antiprotons, photons and neutrinos from LSP annihilation
inside the Galactic Halo.  This ``indirect" detection of DM has
received a lot of attention in recent years following the results of
PAMELA~\cite{Adriani:2008zr} and FERMI~\cite{Abdo:2009zk}.

What are the prospects for indirect detection within our framework?
The WIMP DM candidate in the allowed region of parameter space tends
to be light, i.e $m_{\chi^{0}_1} \lesssim m_W$.  Also, although during
freeze-out, $\chi^{0}_1$ acts like a (pseudo) Dirac
particle,\footnote{In the sense that kinetic energy of $\chi^{0}_1$
during the freeze-out era is large enough to allow $\chi^{0}_1$ and
$\chi^{0}_2$ to co-annihilate.} it acts like a Majorana particle at
present since the kinetic energy of $\chi^{0}_1$ is not enough to
overcome the mass difference between $\chi^{0}_1$ and $\chi^{0}_2$.
Therefore, the dominant co-annihilation contribution to the
cross-section giving rise to the relic-abundance in
Fig.~\ref{fig:relic-parspace} is not operative.  It follows that the
signal for cosmic ray electrons and positrons is significantly below
that seen by PAMELA or FERMI. The PAMELA and FERMI signal must have a
purely astrophysical interpretation within this framework.  A similar
statement holds for cosmic $\gamma$-rays produced by
$\chi^{0}_1\,\chi^{0}_1$ annihilations.  The DM annihilation signal is
too weak to explain the $\gamma$-rays observed by
FERMI~\cite{Fermigamma}, hence those should have a purely
astrophysical explanation as well.

What about cosmic ray neutrinos?  Here, the situation is qualitatively
different for a number of reasons.  As the solar system moves in the
galactic halo, the WIMPs occasionally scatter with the nuclei in the
sun and can lose momentum and become gravitationally bound with it.
For a wide-range of choices of interaction parameters, an equilibrium
can be established between the annihilation and capture rate over the
lifetime of the sun.  Among the various annihilation products of WIMPs
inside the sun, neutrinos are unique since only they can travel to the
earth without significant absorption.  Another important difference
between indirect detection via neutrinos and via other cosmic rays
such as photons or electrons is that the signal depends \emph{not} on
the annihilation cross-section $\langle \sigma\,v\rangle$, but on the
spin-independent (SI) and spin-dependent (SD) scattering
cross-sections with nuclei (H and He in the sun).\footnote{This is true
if the capture and annihilation rates have reached equilibrium, which
holds true in our models.}

Since the neutrino signal depends on both SI and SD cross-sections and
there are very stringent bounds on the SI cross-section from XENON100,
having a large SD cross-section is crucial for a detectable signal.
The effective operator (in four-component notation) and SD scattering
cross-section against nucleons arising from $t$-channel exchange of
$Z$ bosons is \cite{Ellis:2001pa}:
\be
{\cal L}_{\chi_1}^{{\rm SD}}& =& b_q^A\,(\bar{\chi}^{0}_1\gamma^{\mu}\gamma^5\chi^{0}_1)(\bar{q}\gamma_{\mu}\gamma^5 q)~,
\\ [0.5em]
b_q^A &=& -T_q^3\,V_{11}\left(\frac{e^2}{4 s^2_{W} c^2_{W} m_Z^2}\right)~,
\hspace{5mm}
V_{11} = |U_{\tilde{H}_d}|^2-|U_{\tilde{H}_u}|^2~,
\nonumber\\
\sigma_{\chi_1^0}^{{\rm SD}} &\approx& \frac{24}{\pi}\,{G_F^2}\,\frac{m_N^2\,m_{\chi_{1}^0}^2}{(m_N+m_{\chi_{1}^0})^2}\,a_N^2~,
\ee
where using the results in \cite{Ellis:2001pa}, one finds approximately
$a_p \approx 0.705\,V_{11},\,a_n \approx -0.555\,V_{11}$, leading to $\sigma_{\chi_1^0,\,p}^{{\rm SD}} \approx 1.77\times
10^{-37}\,V_{11}^2\,{\rm cm^2}$.  Since $\chi^{0}_1$ in the allowed
region has a non-negligible Higgsino component which couples to the
$Z$, $V_{11}$ is an ${\cal O}(1)$ number giving rise to
$\sigma_{\chi^{0}_1,\,p}^{{\rm SD}} \gtrsim 10^{-39}\,{\rm cm^2}$.
The typical SD cross-section in the allowed region is much larger than
the SI cross-section ($ \sigma_{\chi^{0}_1,\,p}^{{\rm SI}}\lesssim
10^{-44}\,{\rm cm^2}$) because the latter is dominated by Higgs
exchange.  Relative to the SD-cross section, the SI contribution is
suppressed by two powers of the small effective coupling of the Higgs
to the nucleon, $g_{\rm hNN} \sim 10^{-3}$.  Since the LSP mass within
the framework is expected to be $\lesssim m_W$, the relevant
constraints come from Super-K \cite{Desai:2004pq}, which can be
satisfied.  Also, within the framework the channels leading to copious
neutrino production ($W^+W^-,\,\tau^+\tau^-,\,t\bar{t}$)
~\cite{Barger:2011em} are not expected to be significant for the same
reason.  Following ~\cite{Halzen:2009vu}, and the projected
sensitivity of the IceCube/DeepCore experiment~\cite{Wiebusch:2009jf},
however, it is expected that this class of models should be observable
in the near future in neutrino telescopes, if the LSP forms an ${\cal
O}(1)$ fraction of DM. As for direct detection, if the LSP is a
negligible fraction of DM, then its detectability crucially depends on
the combination which appears in the capture rate: $(\rho^{{\rm
LSP}}_{{\rm local}}\,\sigma^{{\rm
SD}}_{\chi_1^0,\,H})/m_{\chi_{1}^0}^2$.

Until now we have focused primarily on the LEP and relic-abundance
constraints, as well as on the signals in direct and indirect
detection experiments.  We will briefly discuss collider and
gravitational wave signals for our scenario in Section~\ref{other},
after considering the issue of complex phases in regards to EWBG and
EDM constraints in Section~\ref{phases}.

\section{A Benchmark Example}
\label{benchmarks}

In this section we perform a numerical analysis of the EWPT and DM
properties, without some of the approximations used in the previous
sections.  For instance, we fully take into account small
$R$-violating terms ($B\mu$, $M_{1}$ and $M_{2}$, though not
$A$-terms), and various small \vevs.  We also include the effects of
the temperature-dependent singlet terms [i.e.~those proportional to
$c_{S^2}$ and $c_{S}$ in Eq.~(\ref{cS})].  However, we do not perform
a full effective potential analysis, restricting rather to the thermal
mass contributions at leading order.  We also do not include
CP-violating phases (see the next section for further discussion on
this point).

For a given choice of parameters, we first minimize numerically the
potential at zero temperature, and determine the spectrum of Higgses,
neutralinos and charginos, as well as their composition.  Only the
dimensionless ratios of the dimensionful input parameters are
meaningful, since we can always rescale them a posteriori to normalize
to $v = 174~{\rm GeV}$.  We then consider the finite-temperature
potential to find the critical temperature.  In order to do so, we
increase $T$, and look for all real extrema of the potential (which is
just polynomial in the approximation we are using), identifying
EW-preserving (where only $S$ has a \vev) and violating minima.  We
then compare the potential energies associated with these minima in
order to identify the global one, for the given $T$.  By iteration, we
can then identify the critical temperature, defined by degeneracy of
the EW symmetry preserving and violating minima.  Note also that since
we can take all squark and slepton soft masses to be positive and
large (unlike in the MSSM, they do not play a role whatsoever in
making the transition first-order), we do not need to worry about
color- or charge-violating minima.

As an example, consider the set of parameters (in GeV units where
relevant) given by
\begin{center}
\begin{tabular}{|c|c|c||c|c|c|c|}
\hline
\rule{0mm}{4.5mm}
$m^{2}_{H_{u}}$ & $m^{2}_{H_{d}}$ & $b$ & 
$\lambda_{s}$ & $t_{s}$ & $B_{s}$ & $m^{2}_{s}$  
\\ [0.3em] 
\hline
\rule{0mm}{4.5mm}
$-(100)^{2}$ & $(100)^{2}$ & $(20)^{2}$ & 
$0.8$ & $(111)^{3}$ & $-(100)^{2}$ & $(125)^{2}$ \\ [0.3em] 
\hline
\end{tabular}
\rule{0mm}{4.5mm}
\begin{tabular}{|c|c|c||c|c|c|c|}
\hline
\rule{0mm}{4.5mm}
$\lambda_{T}$& $B_{T}$ & $m^{2}_{t}$ & 
$M_{D_{1}}$ & $M_{D_{2}}$ & $M_{1}$ & $M_{2}$  
\\ [0.3em] 
\hline
\rule{0mm}{4.5mm}
$1$ & $(300)^{2}$ & $(2000)^{2}$ & $60$& $-110$ & $7.5$& $16$ \\ [0.3em] 
\hline
\end{tabular}
\end{center}
We also take $g \approx 0.65$, $g' \approx 0.35$ and $y_{t} \approx
1$, which results in $c_{\phi} \approx 0.92$, $c_{S^{2}} \approx 0.24$
and $c_{S} \approx -2.78~{\rm GeV}$, from Eqs.~(\ref{cphi}) and
(\ref{cS}).  For the zero-temperature radiative corrections we use
$M_{\rm SUSY} = 2~{\rm TeV}$.  The zero-temperature Higgs \vevs~are
given by $v_{u} \approx 173.99~{\rm GeV}$, $v_{s} \approx 62~{\rm
GeV}$, $v_{d} \approx 1.2~{\rm GeV}$, $v_{T} \approx -7\times
10^{-3}~{\rm GeV}$.  Note that $v_{u}/v_{d} \approx 150$, illustrating
that in this framework ``$\tan\beta$'' can be much larger than in the
MSSM context (see next section).  However, this ratio can easily be
changed by changing $b$, which barely has an effect on the main
physical properties of the model.  Also, the triplet \vev~is
sufficiently small to be irrelevant from the point of view of EW
precision constraints.

The spectrum of CP-even ($m_{H_{i}}$), CP-odd ($m_{A_{i}}$) and
charged ($m_{H^{\pm}_{i}}$) Higgses, in GeV, is
\begin{center}
\begin{tabular}{|c|c|c|c||c|c|c||c|c|c|}
\hline
\rule{0mm}{4.5mm}
$m_{H_{1}}$ & $m_{H_{2}}$ & $m_{H_{3}}$ & $m_{H_{4}}$ & 
$m_{A_{1}}$ & $m_{A_{2}}$ & $m_{A_{3}}$ & 
$m_{H^{\pm}_{1}}$ & $m_{H^{\pm}_{2}}$ & $m_{H^{\pm}_{3}}$  
\\ [0.3em] 
\hline
\rule{0mm}{4.5mm}
$116$ & $184$ & $245$ & $2060$ & 
$234$ & $245$ & $1960$ & 
$129$ & $1960$ & $2060$ \\ [0.3em] 
\hline
\end{tabular}
\end{center}
while the neutralino and chargino spectra are given by
\begin{center}
\begin{tabular}{|c|c|c|c|c|c||c|c|c|}
\hline
\rule{0mm}{4.5mm}
$m_{\chi^{0}_{1}}$ & $m_{\chi^{0}_{2}}$ & $m_{\chi^{0}_{3}}$ & $m_{\chi^{0}_{4}}$ & $m_{\chi^{0}_{5}}$ & $m_{\chi^{0}_{6}}$ & 
$m_{\chi^{\pm}_{1}}$ & $m_{\chi^{\pm}_{2}}$ & $m_{\chi^{\pm}_{3}}$  
\\ [0.3em] 
\hline
\rule{0mm}{4.5mm}
$63.2$ & $70.7$ & $107$ & $120$ & 
$241$ & $244$ & 
$107$ & $127$ & $270$ \\ [0.3em] 
\hline
\end{tabular}
\end{center}
It also of interest to note the composition of the two lightest neutral
CP-even Higgses:
\be
H_{1} &\sim& 0.88 \, h^{0}_{u} - 0.003 \, h^{0}_{d} + 0.48 \, s - 0.003 \, T^{3}_{R}~,
\\
H_{2} &\sim& 0.47 \, h^{0}_{u} - 0.008 \, h^{0}_{d} - 0.88 \, s + 0.005 \, T^{3}_{R}~,
\ee
and of the LSP:
\be
\chi^{0}_{1} &\sim& 0.67 \, \tilde{b} + 0.12 \, \tilde{w}^{3} + 0.05 \, \tilde{H}^{0}_{d} + 0.35 \, \tilde{T}^{3} - 0.54 \, \tilde{S} - 0.35 \, \tilde{H}^{0}_{u}~.
\ee
The content of $\chi^{0}_{2}$ is very similar due to the 
pseudo-Dirac nature of the neutralinos.  For this parameter point, we
find that the relic density is $\Omega_{\chi^{0}_{1}} h^{2} \approx
0.11$, in accord with the WMAP constraint~\cite{Komatsu:2010fb}, while
the LSP spin-independent cross-section is given by
$\sigma_{\chi^{0}_{1}N \to \chi^{0}_{1}N} \approx 4.5 \times
10^{-45}~{\rm cm}^{2}$, somewhat below the current XENON100
limit~\cite{Aprile:2011hi}.

Turning now to the finite-temperature analysis, we find that the
critical temperature is $T_{c} \approx 71.0~{\rm GeV}$, while
$v_{c}/T_{c} \approx 1.34$, exhibiting a strongly first-order EWPT,
even when taking into account the points made in~\cite{Patel:2011th}.
For comparison, the analytic formulas given in
Section~\ref{pot-finite} give $T^{\rm analytic}_{c} \approx 62.5~{\rm
GeV}$, while $(v_{c}/T_{c})^{\rm analytic} \approx 1.86$.  The
difference arises from the temperature-dependent singlet terms
proportional to $c_{S^{2}}$ and $c_{S}$.  Nevertheless, we see that
the physics is correctly captured by the simplified analysis.  These
results will shift slightly under a more detailed full effective
potential analysis, but we do not expect that the conclusions will
radically change.

\section{\cancel{CP} Phases - EWBG and EDM's}
\label{phases}

In Sections~\ref{Potential} and \ref{pot-finite} we studied the scalar
potential in the case that all the relevant parameters are real.  This
allows for a transparent understanding of the mechanism behind a
strongly first-order phase transition, and its connection to the DM
sector, that is typical in the scenarios we consider, as discussed in
the previous sections.  However, the production of a baryon asymmetry
during the EWPT requires CP-violating phases.  A complete treatment of
this question, namely a study of the CP-violating sources that enter
in the transport equations, their diffusion in front of the bubble
wall, the communication of CP-violation to the LH quark sector and the
subsequent processing by sphalerons, is beyond the scope of this work.
Nevertheless, we will make a few remarks in connection to the
production of the BAU that suggest that a large enough baryon
asymmetry can be induced, while being easily consistent with the
currently null EDM searches.

To be definite, we will frame our discussion in the context of
``approach IV'' described in Section~\ref{features}, and discussed in
more detail in Appendix~\ref{IV}.  We consider initially the exact
$U(1)_{R}$ symmetric limit.  The model contains four Higgs doublets
($H_{u}$, $H_{d}$, $H'_{u}$, $H'_{d}$, of which only $H_{u}$ and
$H'_{d}$ acquire non-zero \vevs), a SM singlet $S$, a $SU(2)_{L}$
triplet $T$, and a $SU(3)_{C}$ octet $O$.  The superpotential and the
soft breaking terms are given by Eqs.~(\ref{WIV}) and (\ref{softIV})
of Appendix~\ref{IV}, to which one adds the ``supersoft'' operators
[see Eq.~(\ref{supersoft})] $i M_{D_{a}} \lambda_{a}
\tilde{\Sigma}_{a} + \sqrt{2} M_{D_{a}} D_{a} \Sigma_{a} + {\rm
h.c.}$, where $\lambda_{1} = \tilde{B}$, $\lambda_{2} = \tilde{W}$,
$\lambda_{3} = \tilde{g}$, $\Sigma_{1} = S$, $\Sigma_{2} = T$ and
$\Sigma_{3} = O$, and the $D_{a}$ are the auxiliary fields in the SM
vector superfields.  For simplicity, we have focused on a limit where
$H'_{u}$ and $H'_{d}$ are somewhat heavier than the weak scale (due to
a large $\mu'$-term) and can effectively be integrated
out.\footnote{Such a limit need not be essential to reach our
conclusions: for instance, even if the primed Higgses have masses at
the EW scale, as long as their \vevs~are somewhat suppressed our
analysis of the phase transition should remain as a reasonable
approximation.  As $\langle H'_{d} \rangle$ becomes larger (note that
$\langle H'_{u} \rangle$ can remain small due to the approximate
$R$-symmetry), the analysis will become more complicated, but we don't
think the fact that the system can easily display a strongly
first-order EWPT will change.  Similarly, a connection to DM should
remain, as well as the suppression of EDM's, as discussed below.}
However, we will keep them below initially in order to count all
possible phases more transparently.

The physical phases in the model, associated to the Higgs sector, are
given by
\be
&&
{\rm Arg}\left( t_{s} M_{D_{1}}^{\star} \right)~,
~~~
{\rm Arg}\left( B_{s} M_{D_{1}}^{\star 2} \right)~,
~~~
{\rm Arg}\left( B_{T} M_{D_{2}}^{\star 2} \right)~,
~~~
{\rm Arg}\left( B_{O} M_{D_{3}}^{\star 2} \right)~,
\\ [0.5em]
&&
{\rm Arg}\left( \lambda_{S} \mu^{\star} \right)~,
~~~
{\rm Arg}\left( \lambda_{T} \mu^{\star} \right)~,
~~~
{\rm Arg}\left( \lambda'_{s} \mu^{\prime \star} b^{\prime} \right)~,
~~~
{\rm Arg}\left( \lambda'_{T} \mu^{\prime \star} b^{\prime} \right)~.
\ee
It is convenient to choose $M_{D_{a}}$ to be real (by redefining the
$S$, $T$ and $O$ superfields).  Also, the phase of $b'$ can be
absorbed in $H'_{d}$, for instance, and then we can choose $\mu$ and
$\mu'$ to be real by rephasing the $H_{d}$ and $H'_{u}$ superfields.
Thus, we can choose the physical phases to reside in $\lambda_{s}$,
$\lambda^{\prime}_s$, $\lambda_{T}$, $\lambda^{\prime}_{T}$, $t_{s}$,
$B_{s}$, $B_{T}$ and $B_{O}$.  The octet fields play no role during
the phase transition, so we can set $B_{O}$ aside.  Also, since the
triplet \vev~is required to be small by EW precision constraints, any
possible phase in $B_{T}$ is likely to play only a minor role.  In
addition, given that $\langle H_{d} \rangle = \langle H'_{u} \rangle =
0$ (i.e.~we take their squared soft masses to be sufficiently
positive), and that we can integrate out the heavy $H'_{u}$ and
$H'_{d}$, the Higgs potential can be seen to depend only on
$H_{u}^{\dagger}H_{u}$, $H_{d}^{\dagger} H_{d}$ and $S$ (setting $T
\approx 0$, together with the vanishing of the octet $O$ and
squark/slepton fields).  It follows that non-vanishing phases in the
microscopic parameters can only induce a phase in the singlet \vev,
while all the Higgs doublet \vevs~either vanish, or can be taken to be
real by an $SU(2)_{L}$ transformation.

The phases of $\lambda_{s}$ and $\lambda_{T}$ enter in the chargino
and neutralino mass matrices [see Eqs.~(\ref{neutralino}) and
(\ref{chargino})], as does a possible phase in $\langle S \rangle$
($\lambda'_{s}$ and $\lambda'_{T}$ do not enter, as long as the
$H'_{u}$ and $H'_{d}$ superfields are somewhat heavy).  The $S$
\vev~gives an interesting effect, not present in the MSSM, when ${\rm
Arg}\langle S \rangle$ is spacetime-dependent.  Such a case was
considered in Ref.~\cite{Huber:2006wf} in the context of the related
model of Ref.~\cite{Menon:2004wv} (which does not have the $U(1)_{R}$
symmetry, but whose tree-level potential has the same form as ours).
For the purpose of generating the BAU, the most important source of
CP-violation comes from the chargino sector, at second order in the
gradient expansion (assuming no degeneracy).  In
Ref.~\cite{Huber:2006wf}, it was found that the effect of the
spacetime-dependent singlet complex \vev~easily leads to a significant
baryon-to-entropy ratio $\eta = n_{B}/s \propto \Delta \theta_{s} /
(l_{w} T_{c})$, where $l_{w}$ is the bubble wall thickness (typically
large compared to $1/T \sim 1/T_{c}$), and $\Delta \theta_{s}$ is the
change in the phase of the singlet across the bubble wall.  Thus, the
phases in $t_{s}$ and/or $B_{s}$ can generate a significant baryon
asymmetry within the framework via a chargino source with CP-violation
arising from the \textit{change} in the singlet phase, even if
$\lambda_{s}$ and $\lambda_{T}$ are real.

As emphasized in the Introduction, some level of R-violation is
generically expected once the gravitino mass is generated.  In fact,
we pointed out in Section~\ref{DM} that small Majorana gaugino masses
can play a crucial role in order for the LSP to fully account for
the DM content of the universe.  This suggests a second, qualitatively
different possibility for generating the BAU. If all the phases in the
Higgs potential vanish (as assumed in the analysis of
Section~\ref{pot-finite}), while the CP-violating phases arise only
from the suppressed Majorana gaugino masses, $M_{a}$, one expects that
the CP-violating sources will be proportional to ${\rm
Im}(M_{a})/M_{D_{a}}$.  Such a suppression might be welcome in regions
where a too large baryon asymmetry is produced, as suggested by the
results of~\cite{Huber:2006wf}.  It would be interesting to further
study the detailed aspects of EWBG in the previous scenarios (note
that the model studied in \cite{Huber:2006wf} required sizable gaugino
Majorana masses).

The above mechanisms for generating CP-violating sources during the
EWPT should remain operative in the context of ``approach I''
described in Section~\ref{features}.  However, there can exist
differences between approaches I and IV when it comes to EDMs.  In
order to see this, it will be useful to first comment on
``$\tan\beta$''.  We have assumed that the EW \vev, $v$, is carried
mostly by one Higgs doublet ($H_{u}^0$), and therefore in a sense we
are always considering a ``large $\tan\beta$'' scenario.  As is
well-known, in the MSSM one often finds observables that are
$\tan\beta$ enhanced/suppressed.  In particular, large $\tan\beta$
enhancements arise in the down sector when the \vev~\textit{that gives
rise to the down-type fermion masses} is much smaller than $v$.  Such
enhancements are, however, tied to how these masses are generated, as
illustrated by the two operators in Eq.~(\ref{Ldown}).  If the
down-fermion masses arise from the first operator (as in approach I
with a small $R$-violating $B\mu$-term being responsible for $\bf
m_{d}$ and $\bf m_{e}$), the situation is ``MSSM-like'' in regards to
such enhancements.  However, if the down-fermion masses arise from the
second operator (as in approach I with the Dobrescu-Fox mechanism
\cite{Dobrescu:2010mk}, or as in approach IV), then the $\vev$ of
$H_{d}$ is not related to the down Yukawa matrices, $\bf{y_{d,e}}$.
Instead, the measured fermion masses always relate the Yukawa
couplings to $v_{u} \approx v$, and the source of $\tan\beta$
enhancements disappears.  Notice that, as in approach IV, the down
masses may be ultimately connected to a small \vev~like $v'_{d}$ (see
Appendix~\ref{IV}).  Thus, one can define two different (large)
\vev~ratios, $v_{u}/v_{d}$ and $v_{u}/v'_{d}$.  One may then wonder if
the second ratio can play the role of the MSSM $\tan\beta$, but as we
will see the $R$-symmetry prevents ``$\tan\beta$ enhanced'' terms from
appearing.

The above remarks will be useful when considering the issue of EDM's,
which we will first discuss in the context of approach IV. As
emphasized in~\cite{Kribs:2007ac}, the approximate $U(1)_{R}$ symmetry
leads to a significant relaxation of the constraints from the electron
and neutron EDM's (see~\cite{Pospelov:2005pr} for a review).  The
point is that, at one-loop order, the EDM's induced by a
squark/slepton and gluino/chargino/neutralino loop require LR mass
mixing in the sfermion sector.  In the MSSM and many variants,
including the model of Ref.~\cite{Menon:2004wv}, these LR mixings
arise from $A$-terms and the $\mu$-term, the latter effect being
$\tan\beta$ enhanced in the down-type sector.  In our scenario the
situation is different: the $A$-terms are at least loop-suppressed (if
they arise from anomaly mediation), while the $U(1)_{R}$ symmetry
forbids the ``usual $\mu$-term'', i.e.~a superpotential term coupling
the Higgses responsible for the up and down-type fermion masses (in
our case $H_{u} H_{d}'$).  The ``$\mu$-terms'' allowed by the
$U(1)_{R}$ symmetry, $\mu H_{u} H_{d}$ and $\mu H'_{u} H'_{d}$, do not
contribute to LR sfermion mixing, as long as $H_{d}$ and $H'_{u}$ do
not acquire \vevs.  This holds even in the presence of (small)
Majorana gaugino masses, as required by the DM relic abundance
(discussed in Subsection~\ref{relic}).  Note that such Majorana
gaugino masses will induce, at one-loop order, a
$b_{\slash{\!\!\!\!R}} H_{u} H_{d}$ term, with $b_{\slash{\!\!\!\!R}}
\sim [\alpha_{a} M_{a} \mu/(2\pi)]
\log(\Lambda/\mu)$~\cite{Dobrescu:2010mk} where $\alpha_{a}$ is a SM
fine-structure constant and $\Lambda$ is a UV cutoff.  Such a $b$-term
induces in turn a very small $R$-violating \vev~for $\langle H_{d}
\rangle \sim v_{u} b_{\slash{\!\!\!\!R}}/m^{2}_{H_{d}}$, that
translates into \textit{very small} LR sfermion mixings in the down
sector.  In addition, since the down-type fermions get their masses
mostly from the $H'_{d}$ \vev, instead of $H_{d}$, such LR mixings do
not carry any ``$\tan\beta$'' enhancements.  Thus, we conclude that
there are no constraints from one-loop level EDMs in our scenario,
even if the sfermion masses are light ($\sim 1~{\rm TeV}$).  This can
be contrasted to the case of pseudo-Dirac gauginos without the
$U(1)_{R}$ symmetry, studied in~\cite{Hisano:2006mv}: while the
$M_{a}/M_{D_{a}}$ suppression can be common to the generation of the
BAU and EDM's, in the $U(1)_{R}$ models the additional LR suppression
renders the associated EDM's completely harmless.\footnote{We note
that the pseudo-Dirac nature of the gauginos by itself may be
sufficient to allow for a successful EWBG, while being consistent with
EDM searches, for natural values of the sfermion masses: the singlet
can be responsible for a strongly first-order phase transition, such
that in spite of the $M_{a}/M_{D_{a}}$ suppression of the CP-violating
sources, a sizable BAU can be generated.  At the same time, the 1-loop
EDM's are suppressed by the same $M_{a}/M_{D_{a}}$ factor, allowing
for lighter sfermions than in the MSSM.} Similarly, two-loop
Barr-Zee-type diagrams~\cite{Barr:1990vd}, involving a chargino in the
loop, are also expected to give a contribution to EDM's suppressed by
$M_{a}/M_{D_{a}}$, and are therefore negligible.  As pointed out
in~\cite{Kribs:2007ac}, the leading contribution to EDM's in
$U(1)_{R}$ symmetric scenarios corresponds to the Weinberg three-gluon
operator~\cite{Weinberg:1989dx}, $(w/3) G \cdot \tilde{G} \cdot G$,
which could be observable in the near future if the relevant
CP-violating phases are order-one.  However, these phases are
associated with the gluino/octet sector, and therefore are not closely
connected to the relevant phases for EWBG. Also, if the relevant
CP-violating phases responsible for the BAU arise from the Higgs
sector, as discussed in the first EWBG scenario above (say with
very small/real Majorana gaugino masses), the CP-violation is
communicated to the SM fermion sector only at a high-loop order, hence
it is unconstrained by EDM bounds.

We end this section by commenting on EDM's within ``approach I''.  If
the down-type fermion masses arise at one-loop order as
in~\cite{Dobrescu:2010mk},\footnote{However, note that due to the
suppression in the Majorana gaugino masses, which comes on top of the
one-loop suppression, it may be necessary to take rather large values
for $\mu$ in order to generate the bottom Yukawa
coupling~\cite{Dobrescu:2010mk}.  Thus, this scenario may require some
fine-tuning in the Higgs sector.} then the situation is similar to
approach IV described above, since the down-fermion masses effectively
arise from the second operator in Eq.~(\ref{Ldown}) [see discussion on
$\tan\beta$ above].  If, on the other hand, the down-type fermion
masses arise from a small $H_{d}$ \vev~induced by a $R$-violating
$B\mu$-term, then one can expect large $\tan\beta$ enhancements in the
one-loop induced EDMs, that may compensate the $M_{a}/M_{D_{a}}$
suppression associated with the pseudo-Dirac nature of the gauginos.
A more detailed study is then necessary to estimate the bounds on the
sfermion masses, but one can nevertheless expect an improvement
compared to the situation in the MSSM (see e.g.~\cite{Li:2010ax}).
This conclusion also holds for EDM's induced at 2-loop order.

\section{Other Experimental Signatures}
\label{other}

As mentioned at the end of Subsection~\ref{directDetection}, a
combined set of measurements from a variety of experiments in particle
physics and astrophysics will be needed in order to fully test the
framework and zoom in on the preferred region of parameter space.  In
addition to DM direct and indirect detection, or EDM measurements,
this would include collider physics signals at the LHC and other
future collider experiments, as well as a possible gravity wave signal
arising from the strongly first-order EWPT. Although a comprehensive
study of phenomenological consequences will have some model-dependence
and is beyond the scope of this paper, we will attempt to outline the
characteristic signatures of the framework which depend only on its
crucial features.

\subsection{Collider Signatures}

The framework has a number of interesting collider signatures which
can be probed at the LHC. Some of the signatures arise as a
consequence of the broad features of the framework considered in this
paper, while others arise as general consequences of the $R$ symmetry
and hence share signatures with other $R$-symmetric models
considered in the literature.  We will focus primarily on the former
and only briefly mention the latter, directing the reader to the
relevant papers for reference.

The class of models considered here have a characteristic spectrum.
The lightest degrees of freedom consist of the lightest CP-even Higgs,
and the lightest chargino and neutralino, with the lightest neutralino
being the LSP. They all have masses $\lesssim 120$ GeV. The other
Higgses, charginos and neutralinos generically are heavier, spanning a
large range between around 150 to several hundred GeV. However, in
order to satisfy the constraints on the $T$-parameter, the triplet
scalars have to be quite heavy, in the multi-TeV range.  What about
squarks, sleptons and gluinos?  Since the physics underlying
electroweak baryogenesis and DM in this work is essentially
independent of them, and given the lack of significant constraints
from EDM bounds, the masses of these particles are quite
unconstrained, and can range from a few hundred GeV to multi-TeV.

The lightest CP-even Higgs in the allowed region of parameter space
generically has a non-trivial singlet component.  Therefore, the LEP
constraints on the Higgs mass are somewhat relaxed, the exact amount
depending on the parameters in detail.  Although there exist
additional tree-level contributions to the Higgs quartic in a model
with the coupling $\lambda_s S H_u H_d$, the extra contribution
vanishes in the $v_d \rightarrow 0$ limit, implying that radiative
corrections are important in raising the Higgs mass.  It is more
challenging to discover the lightest CP-even Higgs in this class of
models than in the SM or in the MSSM, due to two main reasons.  First,
as mentioned above, the Higgs has a sizeable singlet component which
reduces its coupling to gauge bosons and quarks.  Second, the decay
mode $h \rightarrow \chi^{0}_1\,\chi^{0}_1$ is generically available,
allowing the Higgs to decay invisibly with an appreciable branching
ratio.\footnote{The lightest Higgs is expected to be below the weak
boson production threshold.} In such a case, vector boson fusion is
expected to provide the most effective search channel at the
LHC~\cite{Choudhury:1993hv}.  On the other hand, the orthogonal, and
heavier combination of $h^{0}_{u}$ and $s$ (call it $H$), can have a
sizable coupling to gauge bosons and to the top quark, and be in a
range where its decay into W/Z pairs has a sizable branching fraction.
Such a state may be looked for in $gg \to H \to WW/ZZ$.  It hardly
needs to be emphasized that Higgs physics in this framework is
extremely rich and interesting, and should be thoroughly investigated.

The signatures related to the lightest neutralino and chargino are
also very important since they are intricately tied to the DM physics.
The lightest chargino in the allowed region tends to be close to the
LEP bound; hence should be discovered at the LHC. What about the LSP?
The pseudo-Dirac nature of the LSP gives rise to two quasi-degenerate
states $\chi^{0}_1$ and $\chi^{0}_2$ split by a small amount $\delta
m$, causing $\chi^{0}_2$ to decay to $\chi^{0}_1$.  Since $\delta m
\ll m_{\chi^{0}_1}$, the decay length $L$ could be large in some
cases, leading to a displaced vertex.  However,
since $L \sim \delta m^{-5}$, the decay length is quite sensitive to
$\delta m$ and could span a large range.

\begin{figure}[t]
\centerline{ \hspace*{-0.5cm}
\includegraphics[width=0.55 \textwidth]{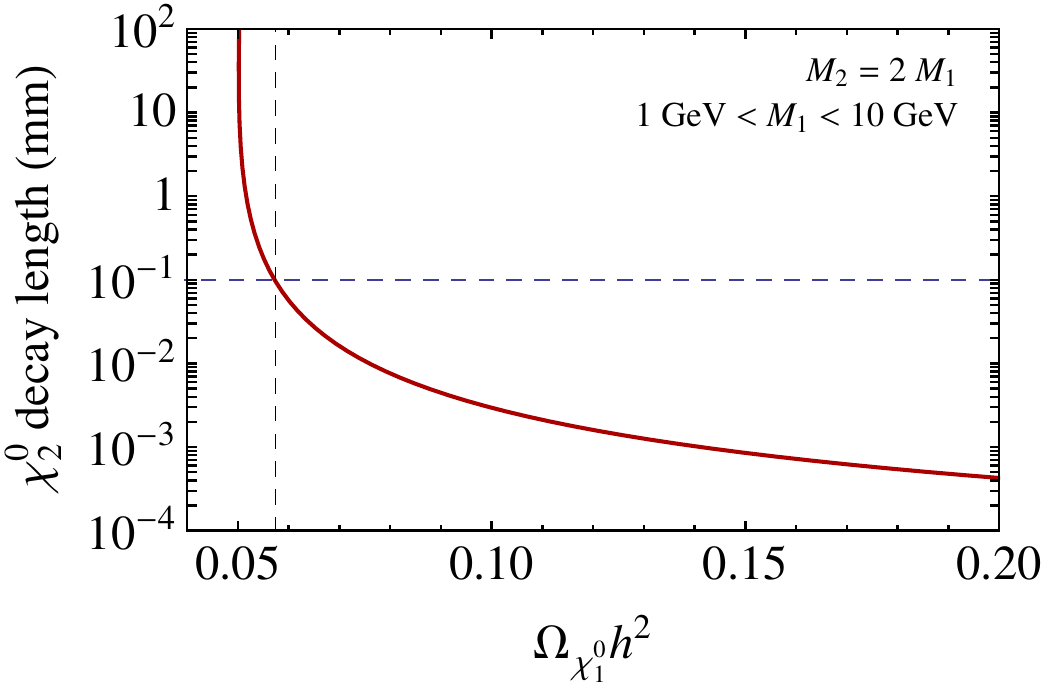}
} 
\caption{\footnotesize{Correlation between the $\chi^{0}_{2}$ rest
frame decay length and the $\chi^{0}_{1}$ relic abundance in the model
of Section~\ref{benchmarks}, as the $U(1)_{Y}$ Majorana gaugino mass,
$M_{1}$, is varied in the range $[1,10]$ GeV. The dashed horizontal line 
indicates approximately the minimum decay length that is measurable.}}
\label{fig:DecayLength}
\end{figure}
Note that $\chi^{0}_2$ decays to $\chi^{0}_1$ and SM fermion pairs
($\bar{f}f$) through an operator which is precisely the same as the
one relevant for computing the relic abundance of the LSP, {\it
viz.}~$\chi^{0}_1$ and $\chi^{0}_2$ co-annihilation to $\bar{f}f$
through $Z$-exchange!  As observed in~\cite{DeSimone:2010tf}, if the
$\chi^{0}_{2}$ decay length is measurable, it allows for a non-trivial
``measurement'' of the relic density, solely based on collider
data, that can then be compared to the cosmological
observations.  The point is that the $\chi^{0}_2$ decay width, given
by
\be
\Gamma(\chi^{0}_{2} \to \chi^{0}_{1} f\bar{f}) &\approx& 
\frac{C^{\prime 4}}{120 \pi^3} \frac{\delta m^5}{M_{Z}^4}~,
\ee
depends only on $\delta m = m_{\chi^{0}_{2}} - m_{\chi^{0}_{1}}$ and
on a ``coupling constant'' $C'$ which can be effectively identified
with the constant $C$ appearing in the annihilation cross-section,
Eq.~(\ref{C}).  The only difference is that the bottom quark
contributes to $C$, but not to $C'$, which results in $C \approx 1.18
C'$.  One can then see that there is a correlation between
$\Omega_{\chi^{0}_{1}}h^{2}$ and the $\chi^{0}_{2}$ decay length
(given the two measurable quantities $m_{\chi^{0}_{1}}$ and $\delta
m$) that is independent of the detailed composition of the LSP. A
difference between our situation and that of
Ref.~\cite{DeSimone:2010tf} is that in our case the annihilation
proceeds in the vicinity of the $Z$ resonance, while
in~\cite{DeSimone:2010tf} both the annihilation and the decay width of
the heavier pseudo-Dirac state were assumed to proceed via a very
heavy state, leading to an effective contact interaction.  As a
result, we find that it is possible for the decay length to be
macroscopic for pseudo-Dirac mass splittings that result in an
\textit{order one} relic density.

As an example, we show in Fig.~\ref{fig:DecayLength} the
$\chi^{0}_{2}$ decay length versus $\Omega_{\chi^{0}_{1}}h^{2}$ in the
benchmark example of Section~\ref{benchmarks}.  We only vary the
Majorana gaugino masses, imposing for concreteness the
``unified-like'' relation $M_{2} = 2 M_{1}$, and letting $1~{\rm GeV}
< M_{1} < 10~{\rm GeV}$.  In this range, the mass splitting between
the two lightest neutralino states (with masses of about $60~{\rm
GeV}$) varies as $0.85~{\rm GeV} \lesssim \delta m \lesssim 10~{\rm
GeV}$.  We then find that the $\chi^{0}_{2}$ decay length in the rest
frame is in the range $10~{\rm cm} \gtrsim L \gtrsim 0.5~{\rm \mu m}$,
and the relic density satisfies $0.05 \lesssim
\Omega_{\chi^{0}_{1}}h^{2} \lesssim 0.19$.  Other properties, such as
the strength of the phase transition, are not sensitive to the
Majorana masses.  Also, in the full range above, the spin-independent
cross-section is consistent with current bounds.  Thus, we see that
${\cal O}(1)$ relic abundances can be associated with observable
($\gtrsim 10^{-1}$ mm~\cite{Aad:2009wy,TDRC}) displaced vertices.
Note that if the LSP is only one of several components of the total DM
abundance, the above method (provided $L$ is measurable) can give a
very useful handle in determining its abundance.  However, it is also
possible that $\chi^{0}_{2}$ is stable on collider time scales, in
which case it could not be distinguished from the real LSP in a
collider environment.

Before moving on to other signatures, we briefly mention some collider
signatures which depend only on the existence of an approximate
$R$-symmetry and are shared with other $R$-symmetric models in the
literature.  In the pure $R$-symmetric limit, it can be shown that
cross-sections for many scattering processes, such as $qq' \rightarrow
\{\tilde{q}_L\tilde{q}'_L, \tilde{q}_R\tilde{q}'_R\}$, $q\bar{q}'
\rightarrow \{\tilde{q}_L\tilde{q}'_R, \tilde{q}_R\tilde{q}'_L\}$,
$q\,g \rightarrow \{\tilde{q}_L\tilde{g}_D, \tilde{q}_R\tilde{g}_D\}$,
and $g\,g \rightarrow \{\tilde{g}_D\tilde{g}_D,
\tilde{g}^c_D\tilde{g}^c_D\}$, vanish\footnote{$\tilde{g}_D$ stands
for the Dirac gluino.}~\cite{Choi:2008pi}.  Thus, these scattering
processes are suppressed if the $R$-breaking is small.  The
suppression of Majorana masses also implies suppression of same-sign
lepton signals.  In the Higgs sector, the presence of the $R$-charge 2
scalar $H_d$ implies that it can only be pair-produced at colliders.
At the LHC, Drell-Yan production mediated by weak gauge bosons is
likely to be the dominant channel~\cite{Choi:2010an}.  Also, in the
$R$-symmetric limit each $H_d$ decays to \emph{two} neutralinos (with
$R$-charge 1) giving rise to \emph{four} neutralinos for events with
pair-produced $H_d$'s.  Finally, since the $R$-symmetric SSM can lead
to a natural suppression of FCNCs even for ${\cal O}(1)$ flavor
violating soft masses, the flavor violations in the soft mass-matrices
can show up in variety of ways at the LHC \cite{Kribs:2007ac}, which
is worthy of significant study.

\subsection{Gravitational Waves}
\label{GravWaves}

A strongly first-order EWPT proceeds by formation of bubbles of the
broken phase which expand into the unbroken phase.  Near the end of
the phase transition, bubbles of the broken phase collide with each
other, breaking the spherical symmetry and leading to production of
gravitational waves~\cite{Caprini:2006jb}-\cite{Espinosa:2010hh}.
Thus, a strongly first-order EWPT can give rise to gravitational
waves, which are in principle observable by space-based
interferometers like LISA or BBO.

A detailed study of the feasibility of observing a gravitational wave
signal arising from the EWPT within our framework is beyond the scope
of this paper, and we will limit ourselves to a few remarks.  As
explained in Section~\ref{pot-finite}, the structure of the finite
temperature potential within the framework is qualitatively different
than in the MSSM or the SM, which naturally allows the possibility of
a strongly first-order EWPT. The crucial feature is the presence of a
barrier due to a negative effective quartic term in the effective
potential, which is balanced by a ``tower of operators" arising at
higher order (see Eqs.~(\ref{VeffSmallphi})-(\ref{lambdaeff}) and the
discussion below).  It turns out that a stronger phase transition
proceeds at a lower temperature and creates larger bubbles, shifting
the peak of the gravity wave spectrum to lower frequencies, below the
best sensitivity range of satellite experiments like LISA and BBO.
However, as explained in \cite{Huber:2008hg}, since the gravity wave
spectrum has a much milder fall-off with frequency ($\sim f^{-1.0}$)
than thought earlier \cite{Huber:2007vva}, it is expected that BBO
would be sensitive to gravity waves from the EWPT. Also, as pointed
out recently in \cite{No:2011fi}, EWBG and gravity waves depend on
different velocities associated with the expanding bubble wall.  While
gravity waves depend on the wall velocity $V_w$ as a whole,
electroweak baryogenesis is sensitive to the relative velocity $v_+$
between the wall and the plasma in front, which is in general smaller
than $V_w$, the difference being more pronounced for stronger phase
transitions.  Thus, it may be possible to have an observable gravity
wave signal from BBO or LISA from a strong electroweak phase
transition responsible for electroweak baryogenesis \cite{No:2011fi}.

\section{Summary and Conclusions}
\label{conclusions}

In this work we have considered a supersymmetric extension of the SM
that have an approximate $U(1)_{R}$ symmetry.  We point out that 
this class of models can naturally lead to a strongly first-order
electroweak phase transition, and possibly to the successful
production of the observed baryon asymmetry, without the tensions that
plague the MSSM. We have further pointed out that there is a close
connection between the EWPT and the properties of DM
(see~\cite{Cirigliano:2006dg} for a study of the DM/EWBG connection in
the context of the MSSM).

The basic observations derive from the (pseudo) Dirac nature of
gauginos in $R$-symmetric models.  This is motivated by the fact
that Majorana gaugino masses are forbidden by the $R$-symmetry, which
leads one to introduce superfields in the adjoint representation of
the SM gauge group.  In particular, the Dirac partner of the bino
arises from a SM singlet superfield, whose scalar component plays an
essential role in leading to a first-order EWPT. Since this is a
tree-level effect, the phase transition is easily strong, even for
Higgs masses comfortably above the Higgs LEP bound, thus creating the
necessary conditions for generating the BAU. We find that the
interesting region of parameter space is also characterized by a
sizable Higgsino component of the lightest neutralino (the LSP), which
results in an appealing DM/EWBG connection.  This fact, together with
the (pseudo) Dirac nature of the LSP, leads to rather interesting DM
physics: at the time of freeze-out, the LSP behaves similarly to a
Dirac fermion, and has a sizable annihilation cross-section into
fermion pairs via Z-exchange.  This cross-section can be modulated by
a small Majorana mass splitting.  On the other hand, for annihilations
today, or in scattering against nuclei, the LSP behaves like a
Majorana fermion, thus being consistent with current constraints.
Nevertheless, one expects that such a DM candidate will be observable
in upcoming direct detection experiments, and possibly also at
neutrino telescopes, such as IceCube/DeepCore.

We have further stressed that, unlike in the MSSM, the CP-violation
that is relevant in the generation of the BAU is typically not
constrained by EDM searches (although EDM signals are possible within
the framework).  We have also commented on the collider prospects,
which are characterized by a rich Higgs sector, and a possibly very
interesting signal associated with the two semi-degenerate lightest
neutralino states - in the form of a visible displaced vertex in
association with missing energy, arising from the decay of $\chi_2^0$
to $\chi_1^0$.  Moreover, the collider measurements of the decay
length $L$, $m_{\chi_1^0}$ and $\delta m$ allows the ``measurement" of
a cosmological observable - the relic abundance of the LSP. Due to the
strength of the EWPT, gravitational wave signals represent another
exciting possibility.

In conclusion, the present class of models represents a well-motivated
extension of the SM (addressing the hierarchy problem, while
significantly alleviating the SUSY flavor and CP problems), that can
also easily lead to a successful generation of the BAU during the
EWPT, while providing a non-standard DM candidate.  In all these
regards, it compares favorably to the MSSM. More detailed studies of
the points made above are worth pursuing and are left for future work.

\section*{Acknowledgments}
The work of EP and PK is supported by the DOE grant DE-FG02-92ER40699.  

\appendix

\section{Expressions including $R$-breaking and $v_T \neq 0$}

Here we collect the relevant expressions when $R$-symmetry breaking
terms by Majorana gaugino masses and the $b \equiv B\mu$ term are
included, and when $v_T \neq 0$.  However, we still assume that
$A$-terms are negligible.  Now, unlike in the analysis in
Section~\ref{Potential}, the $H_d$ degrees of freedom no longer
decouple from the system and a small {\it vev} $v_d$ is induced for
$H_d^0$.  Also, the triplet vev $v_T$ will be small but non-zero.  For
simplicity, we will only consider a vacuum which is CP-preserving,
i.e.~only the real parts of the fields get vevs.  The minimization
conditions are then given by:
\be
m_{H_u}^2&=&b\,t^{-1}_{\beta}+ \left[\frac{(g^2+g'^2)}{4} \,c_{2\beta}- s^2_{\beta} \, \Delta\lambda -(\lambda_s^2+\lambda_T^2)\,c^2_{\beta} \right] v^2
\nonumber\\
&&-\sqrt{2}\,(g' M_{D_1} v_s - g M_{D_2} v_T)
-(\lambda_s v_s+\lambda_T v_T)^2~,
\nonumber\\ [0.5em]
m_{H_d}^2&=&b\,t_{\beta}- \left[\frac{(g^2+g'^2)}{4}\, c_{2\beta} + (\lambda_s^2+\lambda_T^2) s^2_{\beta} \right] v^2
+\sqrt{2}\,(g' M_{D_1} v_s-g M_{D_2} v_T)
\nonumber\\ [0.5em]
&&-(\lambda_s v_s+\lambda_T v_T)^2~,
\nonumber\\ [0.5em]
m_{s_R}^2&=&\frac{\sqrt{2}g' M_{D_1} v^2 c_{2\beta} - 2\,t_s - 2\,\lambda_s\,(\lambda_s\,v_s+\lambda_T\,v_T)\,v^2}{2\,v_s}~,
\nonumber\\ [0.5em]
m_{T_R}^2&=&\frac{-\sqrt{2}g M_{D_2} v^2 c_{2\beta} -2\,\lambda_T\,(\lambda_s\,v_s+\lambda_T\,v_T)\,v^2}{2\,v_T}~.
\ee

We now write the components of the general four-dimensional neutral
CP-even, neutral CP-odd and the charged Higgs scalar mass-squared
matrices.  Since these are symmetric, there are ten independent
components.  For the CP-even scalars in the
($h_u^0,h_d^0,S_{R},T^3_R$) basis, and denoting the mass-squared
matrix as ${\cal M}_{H}^2$, one has:
\be 
\label{H-CPeven}
{\cal M}_{H,11}^2&=& \left[\frac{g^2+g'^2}{4}(1-c_{2\beta})+3s^2_{\beta}\,\Delta\lambda+(\lambda_s^2+\lambda_T^2)c^2_{\beta} \right] v^2 +
\nonumber\\ [0.5em] && 
(\lambda_s v_s+\lambda_T v_T)^2 + \sqrt{2} (g' M_{D_1} v_s-g M_{D_2} v_T) +m^2_{H_u}~,
\nonumber\\
{\cal M}_{H,21}^2&=& -b-\left[\frac{g^2+g'^2}{4}-(\lambda_s^2+\lambda_T^2)\right]s_{2\beta}\,v^2~,
\nonumber\\ [0.5em]
{\cal M}_{H,31}^2&=& \left[\sqrt{2}g'M_{D_1}+2\lambda_s(\lambda_s\,v_s+\lambda_T\,v_T)\right] s_{\beta}\,v~,
\nonumber\\ [0.5em]
{\cal M}_{H,41}^2&=& \left[-\sqrt{2}g M_{D_2}+2\lambda_T (\lambda_s\,v_s+\lambda_T\,v_T \right] s_{\beta}\,v~,
\nonumber
\ee
\be
{\cal M}_{H,22}^2&=&[\frac{g^2+g'^2}{4}(1+2c_{2\beta})+(\lambda_s^2+\lambda_T^2)s^2_{\beta}]\,v^2+(\lambda_s v_s+\lambda_T v_T)^2-
\nonumber\\ [0.5em]
&&\sqrt{2} (g' M_{D_1} v_s - g M_{D_2} v_T)+m^2_{H_d}~,
\nonumber\\
{\cal M}_{H,32}^2&=& \left[-\sqrt{2}g'M_{D_1}+2\lambda_s(\lambda_s\,v_s+\lambda_T\,v_T)\right] c_{\beta}\,v~,
\\ [0.5em]
{\cal M}_{H,42}^2&=& \left[\sqrt{2}g M_{D_2}+2\lambda_T (\lambda_s\,v_s+\lambda_T\,v_T) \right] c_{\beta}\,v~,
\nonumber\\ [0.5em]
{\cal M}_{H,33}^2&=& m_{S_R}^2+\lambda_s^2\,v^2~,
\hspace{4mm}
{\cal M}_{H,43}^2 ~=~ \lambda_s\,\lambda_T\,v_s\,v_T~,
\hspace{4mm}
{\cal M}_{H,44}^2~=~ m_{T_R}^2+\lambda_T^2\,v^2~.
\nonumber
\ee
For the CP-odd scalars in the ($G^0,A^0,S_I,T^3_I$) basis, and
denoting the mass-squared matrix as ${\cal M}_{A}^2$, one has
\be 
\label{H-CPodd}
{\cal M}_{A,11}^2&=&-b \, s_{2\beta}+ \left[\frac{g^2+g'^2}{4}c^2_{2\beta}+s^4_{\beta}\,\Delta\lambda+\frac{1}{2} (\lambda_s^2+\lambda_T^2)s^2_{2\beta} \right] v^2+(\lambda_s\,v_s+\lambda_T\,v_T)^2-\nonumber\\ [0.5em] && 
\sqrt{2} (g' M_{D_1} v_s - g M_{D_2} v_T)c_{2\beta} + s^2_{\beta}\,m^2_{H_u}+c^2_{\beta}\,m^2_{H_d}~,
\nonumber\\ [0.5em]
{\cal M}_{A,21}^2&=&-b \,c_{2\beta}+\left[-\frac{g^2+g'^2}{8}s_{4\beta}+s^3_{\beta} c_{\beta}\,\Delta\lambda+ \frac{1}{4} (\lambda_s^2+\lambda_T^2) s_{4\beta} \right]v^2+
\nonumber\\
&&\left[2\sqrt{2} \, (g' M_{D_1} v_s-g M_{D_2} v_T)+ (m^2_{H_u}-m^2_{H_d})\right] s_{\beta}c_{\beta}~,
\nonumber \\ [0.5em]
{\cal M}_{A,31}^2&=&0~,
\hspace{4mm}
{\cal M}_{A,41}^2 ~=~ 0~,
\\ [0.5em]
{\cal M}_{A,22}^2&=&b\,s_{2\beta}+\left[-\frac{g^2+g'^2}{4}c^2_{2\beta}+\frac{s^2_{2\beta}}{4}\Delta\lambda+(\lambda_s^2+\lambda_T^2)\left(1-\frac{s^2_{2\beta}}{2} \right) \right] v^2+ (\lambda_s v_s+\lambda_T v_T)^2+
\nonumber\\ [0.5em]
&& \sqrt{2} \, (g' M_{D_1} v_s-g M_{D_2} v_T) \, c_{2\beta} + c^2_{\beta}\,m^2_{H_u}+s^2_{\beta}\,m^2_{H_d}~,
\nonumber\\ [0.5em]
{\cal M}_{A,32}^2&=&0~,
\hspace{4mm}
{\cal M}_{A,42}^2~=~0~,
\hspace{4mm}
{\cal M}_{A,33}^2~=~ m_{s_R}^2-4 B_s-4 M_{D_1}^2+\lambda_s^2\,v^2~,
\nonumber\\
{\cal M}_{H,43}^2&=& \lambda_s\,\lambda_T\,v^2~,
\hspace{4mm}
{\cal M}_{A,44}^2~=~ m_{T_R}^2-4 B_T-4 M_{D_2}^2+\lambda_T^2\,v^2~.
\nonumber
\ee
The mass-squared matrix elements for the charged Higgs scalars,
written as ${\cal H}^+{\cal M}_{H^{\pm}}^2 {\cal H}^-$, with ${\cal
H}^+ = \{H_u^+,H_d^+,T_u^+,T_d^+\}$ and ${\cal H}^- =
\{H_u^-,H_d^-,T_u^-,T_d^-\}$, are:

\be\label{H-charged}
{\cal M}_{H^{\pm},11}^2&=&\left[\frac{g^2}{4}-\frac{g'^2}{4}c_{2\beta} + s^2_{\beta}\,\Delta\lambda\right]v^2+(\lambda_s v_s-\lambda_T v_T)^2+\sqrt{2} \, (g' M_{D_1} v_s+g M_{D_2} v_T ) +m^2_{H_u}~,
\nonumber\\
{\cal M}_{H^{\pm},21}^2&=& b+\left(\frac{g^2}{2}-\lambda_s^2+\lambda_T^2\right) s_{\beta} c_{\beta} \, v^2~,
\nonumber\\
{\cal M}_{H^{\pm},31}^2&=&\left[2\,g\,M_{D_2}+\sqrt{2}\,g^2\,v_T- 2\sqrt{2}\,\lambda_T (\lambda_s v_s - \lambda_T v_T) \right] \frac{s_{\beta}}{2}\,v~,
\nonumber\\
{\cal M}_{H^{\pm},41}^2&=&-\left[-2\,g\,M_{D_2}+\sqrt{2}\,g^2\,v_T+2\sqrt{2}\,\lambda_T (\lambda_s v_s + \lambda_T v_T) \right] \frac{s_{\beta}}{2}\,v~,
\nonumber\\
{\cal M}_{H^{\pm},22}^2&=& \left[\frac{g^2}{4}+\frac{g'^2}{4}c_{2\beta} \right] v^2+(\lambda_s v_s-\lambda_T v_T)^2 - \sqrt{2} (g' M_{D_1} v_s + g M_{D_2} v_T) + m^2_{H_d}~,
\\
{\cal M}_{H^{\pm},32}^2&=& \left[2\,g M_{D_2}+\sqrt{2}\,g^2 v_T+2\sqrt{2} \,\lambda_T (\lambda_s v_s + \lambda_T v_T) \right] \frac{c_{\beta}}{2}\,v~,
\nonumber\\
{\cal M}_{H^{\pm},42}^2&=&- \left[-2\,g M_{D_2}+\sqrt{2}\,g^2 v_T-2\sqrt{2}\,\lambda_T (\lambda_s v_s - \lambda_T v_T) \right] \frac{c_{\beta}}{2}\,v~,
\nonumber\\
{\cal M}_{H^{\pm},33}^2&=&  \left(\frac{g^2}{2}c_{2\beta}+2\lambda_T^2\,s^2_{\beta} \right) v^2 + g^2\,v_T^2 + 4\sqrt{2}\,g\,M_{D_2} v_T + m_t^2+2\,M_{D_2}^2~,
\nonumber\\
{\cal M}_{H^{\pm},43}^2&=& -g^2\,v_T^2 +2B_{T} +2M_{D_2}^2~,
\nonumber\\
{\cal M}_{H^{\pm},44}^2&=& \left(-\frac{g^2}{2}c_{2\beta}+2\lambda_T^2\,c^2_{\beta}\right) v^2 + g^2\,v_T^2 - 4\sqrt{2}\,g\,M_{D_2} v_T+ m_t^2+2\,M_{D_2}^2~.
\nonumber
\ee
Similarly, for charginos in the $\{\tilde{T}^+,\tilde{W}^+,\tilde{H}_u^+\,;\,\tilde{T}^-,\tilde{W}^-,\tilde{H}_d^-\}$ basis, one has:
\be 
\label{chargino-full}
{\cal M}_{\chi^{\pm}} &=& \left( \begin{array}{cc} 0 & {\bf X_C} \\[0.5em]
{\bf X_C^T} & 0 
\end{array} \right)~, \nonumber
\ee with
\be
{\bf X_C} &=& \left( \begin{array}{ccc}  0 & M_{D_2}& \sqrt{2}\,c_{\beta}\,\lambda_T\,v \\ [0.5em]
M_{D_2} &  M_{2} & \sqrt{2}\,m_W\,s_{\beta}\\[0.5em]
\sqrt{2}\,s_{\beta}\,\lambda_T\,v&  \sqrt{2}\,m_W\,c_{\beta} &  -v_s\,\lambda_s+v_T\,\lambda_T
\end{array} \right)~.
\ee
Finally, the neutralinos become Majorana in character once $R$-breaking is included. Thus, in the $\{i\tilde{B},i\tilde{W}^0,\tilde{H}_d^0,\tilde{T},\tilde{S},\tilde{H}_u^0\}$ basis, the neutralino mass matrix is given by:
\be
{\cal M}_{\chi^0} &=& \left( \begin{array}{cccccc}  M_1 & 0 & -m_Z\,s_w\,c_{\beta} & 0 & M_{D_1}& m_Z\,s_w\,s_{\beta} \\ [0.5em]
0 & M_2 & m_Z\,c_w\,c_{\beta} & M_{D_2} & 0 & -m_Z\,c_w\,s_{\beta}\\[0.5em]
-m_Z\,s_w\,c_{\beta} & m_Z\,c_w\,c_{\beta} & 0 & s_{\beta}\,\lambda_T\,v& s_{\beta}\,\lambda_s\,v & \lambda_s\,v_s+\lambda_T\,v_T\\[0.5em]
0 & M_{D_2} &s_{\beta}\,\lambda_T\,v & 0 & 0 & c_{\beta}\,\lambda_T\,v\\[0.5em]
M_{D_1} & 0 & s_{\beta}\,\lambda_s\,v & 0 & 0& c_{\beta}\,\lambda_s\,v\\[0.5em]
m_Z\,s_w\,s_{\beta} & -m_Z\,c_w\,s_{\beta} & \lambda_s\,v_s+\lambda_T\,v_T & c_{\beta}\,\lambda_T\,v & c_{\beta}\,\lambda_s\,v & 0
\end{array} \right)~.
\nonumber\\
\ee
%

\section{Approach IV: Details of the Model}
\label{IV}

In this section, we explain in some detail approach IV discussed in
Section \ref{features}.  In particular, in addition to the fields in
the $R$-symmetric SSM, we include a vector-like pair of $SU(2)_L$
doublets $H'_u$ and $H'_d$ with $R[H'_u]=2, R[H'_d]=0$ and
$Y[H'_u]=1/2, Y[H'_d]=-1/2$.  The $R$-charges of all the quark and
lepton superfields are equal to unity, while $R[H_u]=0$ and
$R[H_d]=2$.  This allows the following superpotential consistent with
the gauge and $R$-symmetries:
\be
W &=& \left({\bf y_u}\,Q U^{c}\,H_u  +\mu\,H_u\,H_d+ \lambda_s\,S\,H_u\,H_d + 
\lambda_{T}\,T\,H_u\,H_d\right) +
\nonumber\\
& &  \left({\bf y'_d}\,Q D^{c}\,H'_d  +\mu'\,H'_u\,H'_d+ \lambda'_s\,S\,H'_u\,H'_d + 
\lambda'_T\,T\,H'_u\,H'_d\right)~.
\label{WIV}
\ee
The soft supersymmetry breaking terms in the potential consistent with
the $U(1)_{R}$ symmetry are given by:
\be
\label{softIV}
V_{\rm soft} &=& m^2_{H_u}\,|H_u|^2+m^2_{H_d}\,|H_d|^2+ m^2_{H'_u}\,|H'_u|^2+m^2_{H'_d}\,|H'_d|^2+ (b' \,H_u\,H'_d + {\rm h.c.})+\nonumber\\
& & (t_{s} S + B_{s}\,S^2 + B_{T}\,T^2 + {\rm h.c.}) + \sum_i\,m^2_{Q_i}\,|\tilde{Q}_i|^2 + \sum_i\,m^2_{L_i}\,|\tilde{L}_i|^2~,
\ee
where, for simplicity, we omit the octet scalar terms and a possible
cubic term for the singlet (expected to be small).  We denote all the
squarks by $\tilde{Q}$ and all the sleptons by $\tilde{L}$.  We assume
that all the soft mass parameters
$\{m^2_{H_u},m^2_{H_d},m^2_{H'_u},m^2_{H'_d},m^2_{Q},m^2_{L},b',$
$B_{s}, B_{T}\}$ are parametrically of ${\cal O} (M^2_{\rm soft})$
--close to the TeV scale.  We will also assume that $\mu \sim M_{\rm
soft}$, but that $\mu'$ is larger than $M_{\rm soft}$.
Therefore, the scalar and fermionic components of the $H'_u$ and
$H'_d$ superfields are heavy (with masses $\sim \mu'$) and decouple
from the EW scale physics.

Due to the assumption that $\mu' \gg M_{\rm soft}$, $H'_u$ does not
get a \vev, and neither does $H_{d}$ if $m^{2}_{H_{d}} > 0$.  On the
other hand, the $b'$ term in Eq.~(\ref{softIV}) induces a small
\vev~for $H'_d$ once $H_{u}$ gets a non-zero \vev.  This \vev\, will
then induce masses for down-type fermions:
\be
v'_d &\sim&  \frac{v_u b'}{M^2_{H'_d}}
~~~\implies~~~ {\bf m_d} ~\sim~ {\bf y'_d}\,\frac{v_u b'}{M^2_{H'_d}}~.
\ee 
The above result can also be seen in terms of operators by
integrating out the heavy superfields $H'_u$ and $H'_d$.  Since $\mu'$
is assumed to be larger than $M_{\rm soft}$, it is possible
to integrate out the entire supermultiplet.  Solving the equation of
motion for the superfield ${H'_d}^{\dag}$ gives:
\be
\label{Hd'}
{H'_d}^{\dag} = -\frac{1}{4\,{\mu'}^2}\,({\bf {y'}_d}\,{\cal D}^{\alpha}Q\,{\cal D}_{\alpha} D^{c}+{\bf {y'}_e}\,{\cal D}^{\alpha}L\,{\cal D}_{\alpha} E^{c}) + \cdots
\ee
Substituting the above solution for ${H'_d}^{\dag}$ in the $b'$ term,
written as $\int \!  d^{4}\theta \, (X^{\dagger} X/M^{2}_{\star})
H_{u} H'_{d} + {\rm h.c.}$ gives rise to:
\be
L_{\rm down} &=&  -\int d^4\theta \left({\bf y'_d}\,\frac{{\cal D}_{\alpha}Q\,{\cal D}_{\alpha} D^{c}\,H_u^{\dag}}{4{\mu'}^2}\,\frac{X^{\dag}X}{M_{\star}^2} + \,{\bf y'_e}\,\frac{{\cal D}_{\alpha}L\,{\cal D}_{\alpha} E^{c}\,H_u^{\dag}}{4{\mu'}^2}\,\frac{X^{\dag}X}{M_{\star}^2}\right)~, \nonumber
\ee
and therefore
\be
{\bf m_d} &\sim& {\bf y'_d}\,\frac{v_u\,M^{2}_{\rm soft}}{M^2_{H'_d}}~.
\ee
Note that due to the suppression in $v'_{d}$ from $\mu' \gg M_{\rm
soft}$, the bottom Yukawa coupling may be of order one, corresponding
to a large $\tan\beta \equiv v_{u}/v'_{d}$ scenario.  However, as a
result of the structure of the $U(1)_{R}$ symmetric model, this does
not necessarily lead to the typical large $\tan\beta$ enhancements
that characterize certain observables within the MSSM. An example
occurs for the electron and neutron EDM's, as described in
Section~\ref{phases}.


\end{document}